%Paper: hep-th/9212017
%From: POPE@PHYS.TAMU.EDU
%Date: Wed, 2 Dec 1992 11:15:30 CST

%%%%%%%%%%%%%%%%%%%%%%%%%%%%%%%%%%%%%%%%%%%%%%%%%%%%%%%%%%%
%%%                                                     %%%
%%%     The Low-level Spectrum of the $W_3$ String      %%%
%%%                                                     %%%
%%%   H. Lu, B.E.W. Nilsson, C.N. Pope, K.S. Stelle     %%%
%%%                   P.C. West                         %%%
%%%                                                     %%%
%%%                USE PLAIN TEX                        %%%
%%%                                                     %%%
%%%%%%%%%%%%%%%%%%%%%%%%%%%%%%%%%%%%%%%%%%%%%%%%%%%%%%%%%%%

\def\singlespace{\normalbaselines}
\def\oneandahalfspace{\baselineskip=1.15\normalbaselineskip plus 1pt
\lineskip=2pt\lineskiplimit=1pt}

\def\np{\vfill\eject}
\def\nl{\hfil\break}

\def\nofirstpagenoten{\nopagenumbers\footline={\ifnum\pageno>1\tenrm
\hss\folio\hss\fi}}
\def\nofirstpagenotwelve{\nopagenumbers\footline={\ifnum\pageno>1\twelverm
\hss\folio\hss\fi}}
\def\leaderfill{\leaders\hbox to 1em{\hss.\hss}\hfill}
\def\ft#1#2{{\textstyle{{#1}\over{#2}}}}
\def\frac#1/#2{\leavevmode\kern.1em
\raise.5ex\hbox{\the\scriptfont0 #1}\kern-.1em/\kern-.15em
\lower.25ex\hbox{\the\scriptfont0 #2}}
\def\sfrac#1/#2{\leavevmode\kern.1em
\raise.5ex\hbox{\the\scriptscriptfont0 #1}\kern-.1em/\kern-.15em
\lower.25ex\hbox{\the\scriptscriptfont0 #2}}

  %20 point
                   %17 point
  %14 point
 %17 point
 %14 point
 %14 point
 %14 point

\parindent=20pt
\def\narrow{\advance\leftskip by 40pt \advance\rightskip by 40pt}

\def\AB{\bigskip
        \centerline{\bf ABSTRACT}\medskip\narrow}
\def\nonarrower{\advance\leftskip by -40pt\advance\rightskip by -40pt}
\def\AE{\bigskip\nonarrower}
\def\undertext#1{$\underline{\smash{\hbox{#1}}}$}

\def\boxit#1{\vbox{\hrule\hbox{\vrule\kern3pt
        \vbox{\kern3pt#1\kern3pt}\kern3pt\vrule}\hrule}}

\def\gtorder{\mathrel{\raise.3ex\hbox{$>$}\mkern-14mu
             \lower0.6ex\hbox{$\sim$}}}
\def\ltorder{\mathrel{\raise.3ex\hbox{$<$}|mkern-14mu
             \lower0.6ex\hbox{\sim$}}}
\def\dalemb#1#2{{\vbox{\hrule height .#2pt
        \hbox{\vrule width.#2pt height#1pt \kern#1pt
                \vrule width.#2pt}
        \hrule height.#2pt}}}

\font\fourteentt=cmtt10 scaled \magstep2
\font\fourteenbf=cmbx12 scaled \magstep1
\font\fourteenrm=cmr12 scaled \magstep1
\font\fourteeni=cmmi12 scaled \magstep1
\font\fourteenss=cmss12 scaled \magstep1
\font\fourteensy=cmsy10 scaled \magstep2
\font\fourteensl=cmsl12 scaled \magstep1
\font\fourteenex=cmex10 scaled \magstep2
\font\fourteenit=cmti12 scaled \magstep1
\font\twelvett=cmtt10 scaled \magstep1 \font\twelvebf=cmbx12
\font\twelverm=cmr12 \font\twelvei=cmmi12
\font\twelvess=cmss12 \font\twelvesy=cmsy10 scaled \magstep1
\font\twelvesl=cmsl12 \font\twelveex=cmex10 scaled \magstep1
\font\twelveit=cmti12
\font\tenss=cmss10
 
 \font\ninebf=cmbx7 scaled \magstep1
\font\ninerm=cmr7 scaled \magstep1 \font\ninei=cmmi7 scaled \magstep1
\font\ninesy=cmsy7 scaled \magstep1 
\font\eightrm=cmr7 scaled 1140 
 
\font\sevenbf=cmbx7 \font\sevenrm=cmr7 \font\seveni=cmmi7
\font\sevensy=cmsy7 

\catcode`@=11
\newskip\ttglue
\newfam\ssfam

\def\fourteenpoint{\def\rm{\fam0\fourteenrm}
\textfont0=\fourteenrm \scriptfont0=\tenrm \scriptscriptfont0=\sevenrm
\textfont1=\fourteeni \scriptfont1=\teni \scriptscriptfont1=\seveni
\textfont2=\fourteensy \scriptfont2=\tensy \scriptscriptfont2=\sevensy
\textfont3=\fourteenex \scriptfont3=\fourteenex \scriptscriptfont3=\fourteenex
\def\it{\fam\itfam\fourteenit} \textfont\itfam=\fourteenit
\def\sl{\fam\slfam\fourteensl} \textfont\slfam=\fourteensl
\def\bf{\fam\bffam\fourteenbf} \textfont\bffam=\fourteenbf
\scriptfont\bffam=\tenbf \scriptscriptfont\bffam=\sevenbf
\def\tt{\fam\ttfam\fourteentt} \textfont\ttfam=\fourteentt
\def\ss{\fam\ssfam\fourteenss} \textfont\ssfam=\fourteenss
\tt \ttglue=.5em plus .25em minus .15em
\normalbaselineskip=16pt
\abovedisplayskip=16pt plus 4pt minus 12pt
\belowdisplayskip=16pt plus 4pt minus 12pt
\abovedisplayshortskip=0pt plus 4pt
\belowdisplayshortskip=9pt plus 4pt minus 6pt
\parskip=5pt plus 1.5pt
\setbox\strutbox=\hbox{\vrule height12pt depth5pt width0pt}
\let\sc=\tenrm
\let\big=\fourteenbig \normalbaselines\rm}
\def\fourteenbig#1{{\hbox{$\left#1\vbox to12pt{}\right.\n@space$}}}

\def\twelvepoint{\def\rm{\fam0\twelverm}
\textfont0=\twelverm \scriptfont0=\ninerm \scriptscriptfont0=\sevenrm
\textfont1=\twelvei \scriptfont1=\ninei \scriptscriptfont1=\seveni
\textfont2=\twelvesy \scriptfont2=\ninesy \scriptscriptfont2=\sevensy
\textfont3=\twelveex \scriptfont3=\twelveex \scriptscriptfont3=\twelveex
\def\it{\fam\itfam\twelveit} \textfont\itfam=\twelveit
\def\sl{\fam\slfam\twelvesl} \textfont\slfam=\twelvesl
\def\bf{\fam\bffam\twelvebf} \textfont\bffam=\twelvebf
\scriptfont\bffam=\ninebf \scriptscriptfont\bffam=\sevenbf
\def\tt{\fam\ttfam\twelvett} \textfont\ttfam=\twelvett
\def\ss{\fam\ssfam\twelvess} \textfont\ssfam=\twelvess
\tt \ttglue=.5em plus .25em minus .15em
\normalbaselineskip=14pt
\abovedisplayskip=14pt plus 3pt minus 10pt
\belowdisplayskip=14pt plus 3pt minus 10pt
\abovedisplayshortskip=0pt plus 3pt
\belowdisplayshortskip=8pt plus 3pt minus 5pt
\parskip=3pt plus 1.5pt
\setbox\strutbox=\hbox{\vrule height10pt depth4pt width0pt}
\let\sc=\ninerm
\let\big=\twelvebig \normalbaselines\rm}
\def\twelvebig#1{{\hbox{$\left#1\vbox to10pt{}\right.\n@space$}}}

\def\tenpoint{\def\rm{\fam0\tenrm}
\textfont0=\tenrm \scriptfont0=\sevenrm \scriptscriptfont0=\fiverm
\textfont1=\teni \scriptfont1=\seveni \scriptscriptfont1=\fivei
\textfont2=\tensy \scriptfont2=\sevensy \scriptscriptfont2=\fivesy
\textfont3=\tenex \scriptfont3=\tenex \scriptscriptfont3=\tenex
\def\it{\fam\itfam\tenit} \textfont\itfam=\tenit
\def\sl{\fam\slfam\tensl} \textfont\slfam=\tensl
\def\bf{\fam\bffam\tenbf} \textfont\bffam=\tenbf
\scriptfont\bffam=\sevenbf \scriptscriptfont\bffam=\fivebf
\def\tt{\fam\ttfam\tentt} \textfont\ttfam=\tentt
\def\ss{\fam\ssfam\tenss} \textfont\ssfam=\tenss
\tt \ttglue=.5em plus .25em minus .15em
\normalbaselineskip=12pt
\abovedisplayskip=12pt plus 3pt minus 9pt
\belowdisplayskip=12pt plus 3pt minus 9pt
\abovedisplayshortskip=0pt plus 3pt
\belowdisplayshortskip=7pt plus 3pt minus 4pt
\parskip=0.0pt plus 1.0pt
\setbox\strutbox=\hbox{\vrule height8.5pt depth3.5pt width0pt}
\let\sc=\eightrm
\let\big=\tenbig \normalbaselines\rm}
\def\tenbig#1{{\hbox{$\left#1\vbox to8.5pt{}\right.\n@space$}}}
\let\rawfootnote=\footnote \def\footnote#1#2{{\rm\parskip=0pt\rawfootnote{#1}
{#2\hfill\vrule height 0pt depth 6pt width 0pt}}}

\def\tenfoot{\tenpoint\hskip-\parindent\hskip-.1cm}

\overfullrule=0pt
\twelvepoint
\def\sbullet{\raise.2em\hbox{$\scriptscriptstyle\bullet$}}
\nofirstpagenotwelve
%\magnification=\magstep1%
\hsize=16.5 truecm
%\vsize=23.0 truecm
\baselineskip 15pt
%\parskip 0pt
%\font\ftf=cmr8

\def\undertext#1{$\underline{\smash{\hbox{#1}}}$}

\def\ft#1#2{{\textstyle{{#1}\over{#2}}}}

\def\a{\alpha}

\def\del{\partial}

\def\ket#1{\big| #1 \big\rangle}
\def\bra#1{\big\langle #1 \big|}
\def\braket#1#2{\big\langle #1 \big| #2 \big\rangle}

\def\C{\mkern1mu\raise2.2pt\hbox{$\scriptscriptstyle|$}\mkern-7mu{\rm C}}

\oneandahalfspace
\rightline{CTP TAMU--70/92}
\rightline{G\"oteborg ITP 92-43}
\rightline{Imperial/TP/91-92/22}
\rightline{KCL-TH-92-8}
\rightline{hep-th/9212017}
\rightline{December 1992}

\vskip 2truecm
\centerline{\bf The Low-level Spectrum of the $W_3$ String}
\vskip 1.5truecm
\centerline{H.\ Lu$^{(1)}$, B.E.W.\ Nilsson$^{(2)}$,
C.N.\ Pope$^{(1)}$\footnote{$^*$}{\tenfoot \sl  Supported in
part by the U.S. Department of Energy, under
grant DE-FG05-91ER40633.}, K.S.\ Stelle$^{(3)}$\footnote{$^\$$}{\tenfoot \sl
Supported in part by the Commission of the European Communities under
Contract  SC1*-CT91-0674.} and P.C.\ West$^{(2),(4)}$}

\vskip 1.5truecm
\centerline{\it $^{(1)}$ Center
for Theoretical Physics,
Texas A\&M University,}
\centerline{\it College Station, TX 77843--4242, USA.}
\bigskip

\centerline{\it $^{(2)}$Institute of Theoretical Physics,}
\centerline{\it  Chalmers Institute
of Technology and University of G\"oteborg,}
\centerline{\it S-41296 G\"oteborg, Sweden.}
\bigskip

\centerline{\it $^{(3)}$ The Blackett Laboratory, Imperial College,}
\centerline{\it \phantom{xxxxxxx} Prince Consort Road, London SW7 2BZ, UK.}

\bigskip
\centerline{\it $^{(4)}$ Department of Mathematics, King's College, The
Strand, London, UK.}

\vskip 1.5truecm
\AB\singlespace
  We investigate the spectrum of physical states in the $W_3$ string theory,
up to level 2 for a multi-scalar string, and up to level 4 for the
two-scalar string.  The (open) $W_3$ string has a photon as its only
massless state. By using screening charges to study the null physical states
in the two-scalar $W_3$ string, we are able to learn about the gauge
symmetries of the states in the multi-scalar $W_3$ string.
\AE\oneandahalfspace

\np

\noindent
{\bf 1. Introduction}
\bigskip\bigskip

     String theory arose as an attempt to explain hadronic physics.  The
first stage in this development was to deduce from spin-0 (tachyon)
scattering the scattering amplitudes between any of the infinite number of
particles present in the theory.  In this process of factorisation, an
infinite number of harmonic oscillators were introduced, and it was then
that the Virasoro algebra first made its appearance, in connection with the
physical states contained in the oscillator Fock space.  Our present
understanding of string theory gives a central r\^ole to the
Virasoro algebra.  Indeed, string theories are now constructed out
of conformal blocks in such a way that their central charge add up
to 26 (or 10 in the case of superstrings), so as to cancel the conformal
anomaly of the ghosts (resp.\ superghosts).

     Zamolodchikov introduced into conformal field theory a new two-dimensional
algebra containing a spin two and a spin three current [1].  This is not a Lie
algebra in the usual sense, but nevertheless it has been shown that it
possesses
a BRST charge that is nilpotent [2].  It was natural to suggest [3,4] that this
$W_3$ algebra could be used to construct a string theory.  To obtain a
consistent string theory, one must cancel all local anomalies.  In the work of
[5,6], it was seen that this is the case if the matter carries a
representation of the $W_3$ algebra with central charge $c=100$. Unfortunately,
unlike the case of the Virasoro algebra, one cannot in general simply add two
commuting $W_3$ realisations together to form a third in order to build up a
$c=100$ realisation.

     Realisations of the $W_N$ algebras with arbitrary central charge can be
built from $N-1$ scalar fields by means of a quantum generalisation of the
Miura
transformation [7].  These realisations may be generalised to include
arbitrary numbers of scalar fields [8], thus providing a starting point for the
construction of $W$-string theories.  The original realisation of $W_3$
in [7] involved only two scalar fields; we refer to the corresponding string
as  the two-scalar $W_3$ string.  The generalisation to $D+2$ scalar fields
leads to the ($D+2)$-scalar $W_3$ string.

     In [6,9,10,11], arguments were given towards the clarification of the
$W$-string spectrum. In this paper, we start in section 2 by systematically
finding the physical states in the ghost-vacuum sector of the $W_3$ string
at low-lying levels by solving the physical-state conditions. In particular,
for the two-scalar $W_3$ string we solve for all the levels up to and
including level 4.  For the ($D+2$)-scalar $W_3$ string we solve for all
states up to and including level 2.  In section 3, we review a
characteristic feature of these theories: the inner products of the physical
states in the two-scalar realisation at all levels higher than 0 do not
occur in conjugate pairs with respect to the inner-product
momentum-conservation condition, and so must be treated as zero-norm states.
 The implications of this circumstance become more clear in section 4, where
we show which of the physical states turn out to be BRST-trivial,
finding agreement with the states assigned zero norm from the inner-product
analysis. A general pattern for the null states emerges from our study, and
we conjecture that this holds to all levels.  For multi-scalar $W_3$ string
realisations, the pattern found in the two-scalar string is also expected to
generalise, with all non-null physical states in the ghost-vacuum sector
being constructed solely from excitations not involving creation operators
of the ``frozen'' coordinate that is allowed only discrete momentum values by
the $W_3$ constraints. After reviewing the case of the single-scalar
Virasoro string in section 5, in section 6 we show that all of the physical
states at levels $N>0$ of the two-scalar $W_3$ string, all of them null as
seen earlier, may be written in terms of the action of screening operators
on the vacuum state.

\bigskip\bigskip
\noindent
{\bf 2. Physical States }
\bigskip\bigskip

       The $W_3$ algebra [1] takes the form
$$
\eqalign{
[L_n, L_m]&=(n-m)L_{n+m}+{c\over 12}n(n^2-1)\delta_{n+m,0}\ ,\cr
[L_n, W_m]&=(2n-m)W_{n+m}\ ,\cr
[W_n, W_m]&={16\over{22+5c}}(n-m)\Lambda_{n+m}\cr
&\ \ +(n-m)\Big[\ft1{15}(n+m+2)
(n+m+3)-\ft16(n+2)(m+2)\Big]L_{m+n}\cr
&\ \ +{c\over 360}n(n^2-1)(n^2-4)\delta_{n+m,0}\ ,\cr}
\eqno(2.1)
$$
where
$$
\Lambda_n=\sum_m :L_{n+m}L_{-m}:-\ft1{20}\Big[n^2-4-\ft52\big(1-(-1)^m
\big)\Big] L_n
\eqno(2.2)
$$
and the normal ordering means that $:L_pL_q:=L_qL_p$ if $p > q$.
There exists a well-known two-scalar realisation of the $W_3$ algebra.
However, it was found in [8] that one could also have a many-scalar
realisation.
We define
$$
\eqalignno{
T&=-\ft12 (\partial\phi^{(2)})^2-Q\partial^2\phi^{(2)}+\widetilde T\,
&(2.3)\cr
W&=\ft13 (\partial\phi^{(2)})^3 +
Q\partial\phi^{(2)}\partial^2\phi^{(2)} +\ft13 Q^2
\partial^3\phi^{(2)}+2\partial\phi^{(2)}\widetilde T +Q\partial  \widetilde
T\ , &(2.4)\cr}
$$
where
$$
\widetilde T=-\ft12 \partial X^{\hat\mu}\partial X^{\hat\nu}
\eta_{\hat\mu\hat\nu}- a_{\hat\mu} \partial^2 X^{\hat\mu} \ .\eqno(2.5)
$$
Here the scalar fields $X^{\hat\mu}$ are $X^{\hat\mu}=(X^{\mu}, \phi^{(1)})$
with
$\mu=0,1,2, \ldots,D-1$, and $\eta^{\hat\mu \hat\nu}=(-1,1,1,\ldots,1)$ is the
flat $(D+1)$-dimensional Minkowskian metric. $T$ and $W$ furnish a realisation
of the $W_3$ algebra provided that $a^{\hat\mu}a_{\hat\mu}=
\ft13 Q^2-\ft1{12} D$.  Note that
to get the normalisation of equation (2.1), one should multiply
$W_m$ by $2i\over\sqrt{261}$.  This is unnecessary for
the purposes of this paper, and for simplicity we use the normalisation
in (2.4).  The central charge of such a realisation is
$$
c=2+D+12 Q^2+12  a^{\hat\mu} a_{\hat\mu}\ .\eqno(2.6)
$$
These fields provide a suitable string theory if we cancel the
conformal anomaly.  Since the ghosts contribute $-100$ to the central
charge,we require that $c=100$ in (2.6) and so
$$
Q^2=\ft{49}8\ , \qquad
 a^2\equiv a^{\hat\mu} a_{\hat\mu}=\ft1{12}(\ft{49}2-D)\ . \eqno(2.7)
$$

      A representation of the $W_3$ algebra in the language of Laurent
modes is then given by
$$
L_n=\oint T(z)\, z^{-n-2}\ , \qquad W_m=\oint W(z)\, z^{-n-3}\ ,
\eqno(2.8)
$$
It will prove useful to have oscillator expressions for $L_n$ and
$W_n$.  We define
$$
i\del\phi^{(2)}=\sum_n\a_n^{(2)}z^{-n-1}, \qquad
i\del X^{\hat\mu}=\sum_n\a_n^{\hat\mu} z^{-n-1}\ ,\eqno(2.9)
$$
whereupon
$$
\eqalign{
L_n&=\ft12\sum_p :\a_p^{(2)}\a_{n-p}^{(2)}:-i(n+1) Q \a_n^{(2)}+ \widetilde
L_n\ ,\cr
\widetilde L_n&=\ft12 \sum_p :\a^{\hat\mu}_p\a^{\hat\nu}_{n-p}
\eta_{\hat\mu \hat\nu} :-i(n+1) a_{\hat\mu} \a^{\hat\mu}_n\cr}\eqno(2.10)
$$
and
$$
\eqalign{
W_n&=\ft{i}3\sum_{p,q} :\a_{n-p-q}^{(2)}\a_p^{(2)} \a^{(2)}_q:+ Q \sum_p
(p+1):\a^{(2)}_{n-p}\a^{(2)}_p: -\ft{i}3 Q^2(n+1)(n+2)\a^{(2)}_n\cr
&\quad -2i\sum_p\a^{(2)}_{n-p}
\widetilde L_p - (n+2)\widetilde L_n\ .\cr}\eqno(2.11)
$$
Hermiticity of these generators requires that
$$
\eqalign{
\a^{(2)\dagger}_n&=\a^{(2)}_{-n}\ , \qquad  \qquad \qquad
\a^{\hat\mu\dagger}_n=
\a^{\hat\mu}_{-n}\ , \qquad n\ne 0\ ,\cr
\a^{(2)\dagger}_0&=\a^{(2)}_0-2iQ\ , \qquad  \a^{\hat\mu\dagger}_0=
\a^{\hat\mu}_0-2i a^{\hat\mu}\ .\cr}\eqno(2.12)
$$

     In this paper, we shall focus our attention on physical states that
have the ``standard'' form $\ket{{\rm phys}}=\ket{\psi}\otimes \ket{\rm gh}$,
where $\ket{\psi}$ is built purely from ``matter'' oscillators
acting on a momentum state, and $\ket{\rm gh}$ is the ghost vacuum
state.  In terms of the $b,c$ ghosts for spin 2, and the $\beta,\gamma$
ghosts for spin 3, $\ket{\rm gh}$ takes the form $c_1\gamma_1\gamma_2
\ket{0}$ [9,12].  BRST invariance of the state $\ket{\rm phys}$ then implies
that for these ``standard'' states the physical-state conditions on
$\ket{\psi}$ for the $W_3$ string are [9,10]
$$
\eqalign{
L_n\ket\psi &=0=W_n\ket\psi\ , \qquad n\ge 1\ ,\cr
(L_0-4)\ket\psi &=0=W_0\ket\psi\ .\cr}\eqno(2.13)
$$
The intercept of the $W_0$ generator is zero when $W(z)$ is a primary current
with  respect to the whole energy-momentum tensor $T(z)$.   It suffices to
solve  the $L_0, L_1, L_2$ and $W_0$ constraints, since the remainder then
follow by commutation.  To this end, it is more  convenient to rewrite $W_0$
using the other physical-state conditions.  Consider  a primary state
$\ket\chi$, {\it i.e.}\ one that is annihilated by the positive Fourier
modes of the currents $T(z)$ and $W(z)$, with $L_0\ket\chi=h\ket\chi$ and
$W_0\ket\chi=\omega\ket\chi$.  The relevant part of $W_0$ acting on the
state takes the form
$$
\eqalign{
W_0&={i\over6}\Big\{2{\sum}'_{p,q} :\a^{(2)}_{-p-q}\a^{(2)}_p\a^{(2)}_q:
+6\sum_{p>0}{\sum}'_q:\alpha_{-p}^{(2)}\alpha_q^{(2)}\alpha_{p-q}^{(2)}:
+ f\cr
&\qquad+(\hat\a^{(2)}_0)^2\big[36{\cal N}^{(2)}+12(4-h) \big]
-12\sum_{p>0}\alpha_{-p}^{(2)} \widetilde L_{p}\Big\}\,\cr} \eqno(2.14)
$$
where $f=\hat\alpha_0^{(2)}\big(8(\hat\alpha_0^{(2)})^2 +1\big)$, and $\hat
\a^{(2)}_0=\a^{(2)}_0-iQ$ is an hermitean operator  ((2.14) corrects an
expression given in [10]).   ${\cal
N}^{(2)} $ is the number operator for $\a^{(2)}$ oscillators, {\it i.e.\ }
$$
{\cal N}^{(2)}=\sum_{n=1}^{\infty}\a^{(2)}_{-n}\a^{(2)}_n\ .\eqno (2.15)
$$
The primes on summation symbols in (2.14) means that they contain no
zero-mode terms.

      We can analyse the physical-state conditions at each level
independently.  The level number $N$ of a state is defined
as the eigenvalue of the number operator
$$
{\cal N}^{(2)}+\sum_{n>0}\eta_{
\hat\mu \hat\nu}\a^{\hat\mu}_{-n}\a^{\hat\nu}_n \eqno(2.16)
$$
acting on the state.   A state with no oscillator excitations has level
number $0$.   We shall systematically study the physical states up to and
including level $4$.  However, we shall first make some general remarks on
how to solve the physical-state conditions (2.13).

       A state in the Fock space generated by $\a^{(2)}_{-n}$ and
$\a^{\hat\mu}_{-m}$ is of the form
$$
\varepsilon_{\hat\mu_1\ldots\hat\mu_q}\a^{(2)}_{-n_1}\a^{(2)}_{-n_2}
\cdots\a^{(2)}_{-n_p}\a^{\hat\mu_1}_{-m_1}\a^{\hat\mu_2}_{-m_2}
\cdots\a^{\hat\mu_q}_{-m_q}\ket{p^{\hat\mu},\beta}\ ,\eqno(2.17)
$$
where $\ket{p^{\hat\mu},\beta}$ is a highest-weight state defined by
$$
\eqalign{
\a^{\hat\nu}_0\,\ket{p^{\hat\mu},\beta}&=p^{\hat\nu}\,
\ket{p^{\hat\mu},\beta}\cr
\a^{(2)}_0\,\ket{p^{\hat\mu},\beta}&=\beta\,\ket{p^{\hat\mu},\beta}\cr
\a^{\hat\nu}_n\,\ket{p^{\hat\mu},\beta}&=0=\a^{(2)}_n
\,\ket{p^{\hat\mu},\beta}\qquad n \ge 1\cr}\eqno(2.18)
$$
and we refer to $\varepsilon_{\hat\mu_1\ldots\hat\mu_q}$ as polarisation
tensors of the state.  At level $N$, there are $C^{D+2}_N$ such possible
states, with $C^{D+2}_N$ given by
$$
\prod^{\infty}_{n=1}{1\over{(1-x^n)^{(D+2)}}}=\sum^{\infty}_{N=0}
C^{D+2}_{N} x^N\ .\eqno(2.19)
$$
At level $N$, the $(L_0-4)\ket\psi=0$ condition provides only one
constraint, namely the mass-shell equation
$$
\beta(\beta-2iQ)+p^{\hat\mu}(p_{\hat\mu}-2i a_{\hat\mu})=2(4-N)\ .
\eqno(2.20)
$$
The $L_1\ket\psi=0$ and $L_2\ket\psi=0$ conditions provide $C^{D+2}_{N-1}$ and
$C^{D+2}_{N-2}$ constraint equations (not necessarily independent)
respectively.   The other non-trivial condition, $W_0\ket\psi=0$, provides
$C^{D+2}_N$ constraint equations.  It is easier to study the $W_0$
constraint since $W_0$ acting on $\ket\psi$ will not change the level number
of $\ket\psi$.  Thus the number of equations from this constraint is the same
as the number of polarisation-tensor components. The coefficients of the
polarisation tensors in these equations are functions of the momenta $\beta$
and $p^{\hat\mu}$ of the state on which the oscillators act.  We may write
$W_0\ket\psi=0$ in the generic matrix form
$$
{\cal A} \varepsilon=0\ ,\eqno(2.21)
$$
where $\cal A$ is a $C^{D+2}_N \times C^{D+2}_N$ matrix function of $\beta$
and $p^{\hat\mu}$, since the numbers of equations and polarisation-tensor
components are both equal to $C^{D+2}_N$.  In (2.21), $\varepsilon$ is a
column vector of the $C^{D+2}_N$ polarisation-tensor components.  Clearly we
have non-zero solutions only when $\det {\cal A}=0$.  The operator $W_0$
however, which can be written in the form of equation (2.14), always
preserves or lowers $N^{(2)}$, the level number for $\alpha_n^{(2)}$
oscillators. Therefore if we order the components of the column vector
$\varepsilon$ such that the polarisation-tensor components $\varepsilon_1$
associated with  $\a^{(2)}_n$ only come first, followed by the components
$\varepsilon_2$ for  mixed oscillator terms, and finally the components
$\varepsilon_3$ for $\a^{\hat\mu}_n$ oscillators only at the bottom, then
the matrix $\cal A$ will be of the form
$$
{\cal A}=\left(\matrix{M_1&0&0\cr
                       *&M_2&0\cr
                       *&*&M_3\cr}\right)\ ,\eqno(2.22)
$$
where a $*$ indicates non-zero components whose precise form we do not need
to know.  Consequently, we have $\det{\cal A}=\det M_1\cdot \det M_2\cdot
\det M_3$, and so it will vanish if one or more of the sub-determinants
vanishes.

      Examining the form of $W_0$ in equation (2.14), we find that $M_1$,
$M_2$ and $M_3$ do not depend on $p^{\hat\mu}$ and thus they are functions
of  $\beta$ alone.  This results from the fact that the term
$-2i\sum_{p=1}^{\infty} \widetilde L_{-p}\a^{(2)}_p$ lowers $N^{(2)}$ and so
gives contributions only to the starred areas of ${\cal A}$ in equation
(2.22). Indeed when calculating $\det{\cal A}$, we can drop this term
completely. One can easily see that $M_3$ is of the form
$$
(M_3)_{ij}=f\delta_{ij},\eqno(2.23)
$$
where $f$ is defined below eq. (2.14), and so $\det M_3=(f)^{C^{D+1}_N}$.

         As a consequence we find that the $W_0$ constraint implies that to
have a physical state, it must be that  the determinants of any of
$M_1$, $M_2$ or $M_3$ vanishes.  Each determinant is a polynomial in $\beta$
that can be solved to find the allowed roots.   Depending on which determinant
vanishes, we have three cases:

\bigskip
\noindent\undertext{\rm {\it Case 1}, $\det M_1=0$.}
\bigskip

       A physical state that includes terms that involve {\it only}
$\a^{(2)}_{-n}$ oscillators must have $\det M_1=0$, since $\det M_1\ne 0$
implies that terms with only $\a^{(2)}_{-n}$ oscillators must be absent. The
equation $\det M_1=0$ is polynomial of degree $3C^{1}_N$ in $\beta$.

\bigskip
\noindent\undertext{{\it Case 2}, $\det M_2=0$.}
\bigskip

       These states can have terms generated by $\a^{\hat\mu}_{-n}$ and
$\a^{(2)}_{-m}$, but with no terms that only involve $\a^{(2)}_{-m}$ alone,
unless $\det M_1=0$ also.  $\det M_2=0$ is a polynomial equation of
degree $3(C^{D+2}_N-C^1_N-C^{D+1}_N)$ in $\beta$.

\bigskip
\noindent\undertext{{\it Case 3}, $\det M_3=0$.}
\bigskip

       This implies that $f=0$, {\it i.e.}\ $\hat\beta(8\hat\beta^2+1)=0$.
The three allowed roots are $\beta=iQ, \ft67 iQ, \ft87 iQ$. Such physical
states contain excitations only in the $\hat\mu$ directions and thus can
be written in the form
$$
\ket\psi_{\rm eff}\otimes\ket\beta\ ,\eqno(2.24)
$$
where $\ket\psi_{\rm eff}$ is in the Fock space generated by
$\a^{\hat\mu}_{-n}$ alone and $\ket\beta$ is an eigenstate of $\a^{(2)}_0$
which is annihilated by the number operator ${\cal N}^{(2)}$.   The remaining
$L_n$ physical-state conditions become [9,10,11]
$$
\eqalign{
\widetilde L_n \ket\psi_{\rm eff}&=0\qquad n\ge 1\ ,\cr
(\widetilde L_0-a_{\rm eff})\ket\psi_{\rm eff}&=0\ ,\cr}
\eqno(2.25)
$$
where $a_{\rm eff}=\ft{15}{16}$ if $\beta=iQ$, and $a_{\rm eff}=1$ if
$\beta=\ft67 iQ$ or $\ft87 iQ$.  We refer to states of the form (2.24) as
``ordinary'' states.

      In all three cases, we find that the momentum component
$\beta$ of  physical states can take only a discrete set of values.  This
phenomenon of the freezing of the $\beta$ momentum was first observed in [6].
In the above analysis of $W_0\ket\psi=0$, we have assumed that $\ket\psi$ is a
primary state of conformal dimension $4$.   Thus it remains
to solve the $L_1\ket\psi=L_2\ket\psi=0$ constraints in order to single
out the physical solutions for the $\beta$ momentum from all the possible
roots of $\det{\cal A}=0$.  Given any allowed value of $\beta$,
the values of $p^{\hat\mu}$ are restricted by the mass-shell condition in
(2.20).  $L_1\ket\psi=L_2\ket\psi=0$ can be used to solve the polarisation
tensors of the physical states; in fact such solutions exist only for a
limited set of $\beta$ values out of those allowed by $\det{\cal A}=0$ .  For
the two-scalar $W_3$ string such constraints are much stronger.  This is
analogous to the one-scalar Virasoro string, where the number of physical
states
is significantly reduced.

       We now start systematically to solve the physical state conditions for
the $(D+2)$-scalar $W_3$ string up to level 2, and the two-scalar $W_3$
string up to level 4.

\bigskip\bigskip
\noindent\undertext{\it Level $0$}
\bigskip

       The analysis of level 0 states is straightforward.
Consider a state $\ket{p^{\hat\mu},\beta}$ defined in (2.18), which can also
be viewed as a state of the form given in (2.24).  One can easily see that
all the physical states at this level are given by [6,10]
$$
\ket{p^{\hat\mu}_1,iQ}\ ,\qquad\ket{p^{\hat\mu}_2,\ft67 iQ}\ ,
\qquad\ket{p^{\hat\mu}_2,\ft87 iQ}\ ,\eqno(2.26)
$$
subject to
$$
\eqalign{
p^{\hat\mu}_1(p_1-2i a)_{\hat\mu}&=\ft{15}8\ ,\qquad (\beta=iQ)\ ,\cr
p^{\hat\mu}_2(p_2-2i a)_{\hat\mu}&=2\ ,\qquad (\beta=\ft67iQ,\ft87iQ)\ .}
\eqno(2.27)
$$
For the two-scalar $W_3$ string, there is only one $p^{\hat\mu}$, denoted by
$p$.  In this case equation (2.27) yields six solutions [9];
$$
\eqalign{
&\ket{\ft57 i a,iQ}\ ,\qquad\ket{\ft97 i a,iQ}\ ,\qquad
\ket{\ft67 i a,\ft67iQ}\ , \cr
&\ket{\ft67 i a,\ft87 iQ}\ ,\qquad\ket{\ft87 i a,\ft67 iQ}\ ,
\qquad\ket{\ft87 i a,\ft87 iQ}\ .\cr}\eqno(2.28)
$$

\bigskip
\noindent\undertext{\it Level $1$}
\bigskip

     The most general form of the level-1 states is
$$
(\xi\a^{(2)}_{-1}+\xi_{\hat\mu}\a_{-1}^{\hat\mu})\ket{p^{\hat\mu},\beta}
\equiv{\cal P}_1 \ket{p^{\hat\mu},\beta}\ ,\eqno(2.29)
$$
where ${\cal P}_1$ denotes a generic level-1 excitation operator.  After
solving the physical-state conditions of equation (2.13), ${\cal P}_1$ will
be dependent on $p^{\hat\mu}$ and $\beta$; we shall denote it by
${\cal P}_1(p^{\hat\mu},\beta)$.  Following the method for solving the
physical-state conditions described above, we find that the level-1 physical
states  are [6]:
$$
{\cal P}_1(p^{\hat\mu}_1,\ft{10}7 iQ)\ket{p^{\hat\mu}_1,\ft{10}7 iQ}\ ,\qquad
{\cal P}_1(p^{\hat\mu}_2,\ft{11}7 iQ)\ket{p^{\hat\mu}_2,\ft{11}7 iQ}\ ,
\eqno(2.30)
$$
with
$$
p^{\hat\mu}_1(p_1-2i a)_{\hat\mu}=1\ ,\qquad
p^{\hat\mu}_2(p_2-2i a)_{\hat\mu}=\ft{15}8\ .\eqno(2.31)
$$
The polarisation vectors in ${\cal P}_1$ are given by
$$
\xi^{\hat\mu}={\xi p^{\hat\mu}\over {iQ-3\hat\beta}}\ ,\eqno(2.32)
$$
and so we see that these states are actually scalars.

   In addition, there are the ``ordinary'' (case 3) physical states
$$
\eqalign{
&\left.
{\cal P}_1(p^{\hat\mu}_3,iQ)\ket{p^{\hat\mu}_3,iQ}\right. \qquad\qquad
{\rm with}\qquad
p^{\hat\mu}_3(p_3-2i a)_{\hat\mu}=-\ft18\ ,\cr
&\left.
\eqalign{
&{\cal P}_1(p^{\hat\mu}_4,\ft67 iQ)\ket{p^{\hat\mu}_4,\ft67 iQ}\cr
&{\cal P}_1(p^{\hat\mu}_4,\ft87 iQ)\ket{p^{\hat\mu}_4,\ft87 iQ}\cr}
\right\}\qquad {\rm with}\qquad  p^{\hat\mu}_4(p_4-2i a)_{\hat\mu}=0
\ .\cr}\eqno(2.33)
$$
The polarisation vectors for the states (2.33) obey
$$
\xi=0\ ,\qquad (p_{\hat\mu}-2i a_{\hat\mu})\xi^{\hat\mu}=0\ .\eqno(2.34)
$$

       For the two-scalar $W_3$ string we can determine $p$, and we find at
this level that there are a total of 6 physical states, whose polarisations
${\cal P}_1$ follow from the above discussion.  They are four case-1 states
$$
{\cal P}_1\ket{\ft{12}7 i a,\ft{10}7 iQ}\ ,\qquad
{\cal P}_1\ket{\ft27 i a,\ft{10}7 iQ}\ ,\qquad
{\cal P}_1\ket{\ft97 i a,\ft{11}7 iQ}\ ,\qquad
{\cal P}_1\ket{\ft57 i a,\ft{11}7 iQ}\ ,\eqno(2.35)
$$
together with two ``ordinary'' (case-3) states, for which $p=2i a$, owing
to (2.34), is the only possibility.  This excludes the states with $\beta=i
Q$. Thus the two ordinary states are
$$
{\cal P}_1\ket{2i a,\ft67 iQ}\qquad
{\cal P}_1\ket{2 i a,\ft87 iQ}\ .\eqno(2.36)
$$

\bigskip
\noindent\undertext{\it Level $2$}
\bigskip

      The most general form of level-2 states is given by
$$
{\cal P}_2\ket{p^{\hat\mu},\beta}\equiv
(\varepsilon\a^{(2)}_{-1}\a^{(2)}_{-1}+
\varepsilon_{\hat\mu}\a^{(2)}_{-1}\a^{\hat\mu}_{-1}+
\varepsilon_{\hat\mu\hat\nu}\a^{\hat\mu}_{-1}\a^{\hat\nu}_{-1}+
\xi\a^{(2)}_{-2}+\xi_{\hat\mu}\a^{\hat\mu}_{-2})
\ket{p^{\hat\mu},\beta}\ .\eqno(2.37)
$$
We have the ``ordinary'' (case 3) physical states whose $\beta$ values are,
as for case-3 states at all levels, the same as those of the tachyon states
(2.26), namely
$$
{\cal P}_2\ket{p^{\hat\mu},iQ}\qquad {\rm with}\qquad
p^{\hat\mu}(p_{\hat\mu}-2i a_{\hat\mu})=-\ft{17}8\ ,\eqno(2.38)
$$
and
$$
\eqalign{
&{\cal P}_2\ket{p^{\hat\mu},\ft67 iQ}\cr
&{\cal P}_2\ket{p^{\hat\mu},\ft87 iQ}\cr}\qquad {\rm with}\qquad
p^{\hat\mu}(p_{\hat\mu}-2i a_{\hat\mu})=-2\ .\eqno(2.39)
$$
The polarisation tensors of level-2 ordinary physical states all satisfy
$\varepsilon=\xi=\varepsilon_{\hat\mu}=0$, together with
$$
(p^{\hat\mu}-2i a^{\hat\mu})\varepsilon_{\hat\mu\hat\nu}+\xi_{\hat\nu}=0\ ,
\qquad
\varepsilon_{\hat\mu}^{\phantom{\ }\hat\mu} + 2(p_{\hat\mu}-3i a_{\hat\mu})
\xi^{\hat\mu}=0\ .\eqno(2.40)
$$
      We also have a case-1 physical state
$$
{\cal P}_2\ket{p^{\hat\mu}, \ft{12}7 iQ}\qquad {\rm with} \qquad
p^{\hat\mu}(p_{\hat\mu}-2i a_{\hat\mu})=1\ ,\eqno(2.41)
$$
whose polarisations are given by
$$
\eqalign{
\varepsilon_{\hat\mu}&=
\ft87 iQ \varepsilon p_{\hat\mu}\ ,\qquad\qquad\qquad \xi_{\hat\mu}=-\ft25
\varepsilon (p_{\hat\mu}-i a_{\hat\mu})\ ,\cr
\varepsilon_{\hat\mu\hat\nu}&=\ft15 \varepsilon (\eta_{\hat\mu\hat\nu} -5
p_{\hat\mu} p_{\hat\nu})\ ,\qquad \xi=-\ft27 i Q \varepsilon\ .\cr}
\eqno(2.42)
$$
        There are two case-2 physical states at level 2.  Both have
polarisations restricted by
$$
\eqalign{
\varepsilon=\xi=0,\qquad \varepsilon_{\hat\mu}&=(4iQ-3\beta)\xi_{\hat\mu}
\qquad\varepsilon_{\hat\mu\hat\nu}=\ft12(\xi_{\hat\mu}p_{\hat\nu} +
\xi_{\hat\nu}p_{\hat\mu})\ ,\cr
(p^{\hat\mu}-2i a^{\hat\mu})\varepsilon_{\mu}&=0=
(p^{\hat\mu}-2i a^{\hat\mu})\xi_{\hat\mu}
\ ,\cr}\eqno(2.43)
$$
and can occur only for states with $\beta=\ft{10}7 iQ$ or $\ft{11}7 iQ$,
{\it i.e.\ }
$$
\eqalign{
&{\cal P}_2\ket{p^{\hat\mu}, \ft{10}7 iQ}\qquad {\rm with}\qquad
p^{\hat\mu}(p_{\hat\mu}-2i a_{\hat\mu})=-1\ ,\cr
&{\cal P}_2\ket{p^{\hat\mu}, \ft{11}7 iQ}\qquad {\rm with}\qquad
p^{\hat\mu}(p_{\hat\mu}-2i a_{\hat\mu})=-\ft18\ .\cr}\eqno(2.44)
$$

       For the two-scalar $W_3$ string the number of solutions
reduces significantly.  There are a total of three physical states in this
case; one ordinary (case 3) physical state ${\cal P}_2\ket{\ft{17}7 i a,
iQ}$, and two  other (case 1) physical states ${\cal P}_2\ket{\ft{12}7 i a,
\ft{12}7 iQ}$ and ${\cal P}_2\ket{\ft{2}7 i a, \ft{12}7 iQ}$.

\bigskip
\noindent\undertext{\it Level $3$}
\bigskip

       We carry out the analysis at level 3 only for the two-scalar $W_3$
string.  The general states at this level have $10$ independent excitation
terms.  We may write these as
$$
\eqalign{
{\cal P}_3&=\tau\, \a_{-1}^{(2)}\a_{-1}^{(2)}\a_{-1}^{(2)} + \mu\,
\a_{-1}^{(2)}
\a_{-2}^{(2)} +\sigma\, \a_{-3}^{(2)} +\bar\tau\, \a_{-1}^{(1)} \a_{-1}^{(1)}
\a_{-1}^{(1)} + \bar\mu\, \a_{-1}^{(1)} \a_{-2}^{(1)} +\bar\sigma\,
\a_{-3}^{(1)}\cr
&\quad +\rho\, \a_{-1}^{(2)}\a_{-1}^{(2)}\a_{-1}^{(1)} +\bar\rho\,
\a_{-1}^{(2)}
\a_{-1}^{(1)}\a_{-1}^{(1)} +\kappa\, \a_{-1}^{(2)}\a_{-2}^{(1)}+
\bar\kappa\, \a_{-1}^{(1)}\a_{-2}^{(2)}\ .\cr}\eqno(2.45)
$$
Enforcing the physical-state conditions in (2.13), one finds that
there is  no ``ordinary'' physical state.  This situation is analogous to the
one-scalar  Virasoro string, where level-3 states cannot survive the
restriction of the  physical-state conditions.  However, in the case of the
two-scalar $W_3$  string, there are two case-2 physical states:
$$
{\cal P}_3\ket{\ft{17}7 i a, \ft{11}7 iQ} \qquad {\rm and }\qquad
{\cal P}_3\ket{\ft{18}7 i a, \ft{10}7 iQ}\ ,\eqno(2.46)
$$
where ${\cal P}_3$ is given by (2.45) with
$$
\eqalign{
\tau&=\mu=\sigma=\rho=\bar\kappa=0,\qquad \bar\rho=1\ ,\cr
\bar\tau&={12p\over f},\qquad \bar\mu={12\beta(\beta-2i Q)\over f}, \qquad
\bar\sigma=-{24(p-2i a)\over f},\qquad
\bar\kappa=-(p-2ia)\,\cr}\eqno(2.47)
$$
where $f=\hat\beta(8\hat\beta^2+1)$, as well as two case-1 physical states:
$$
{\cal P}_3\ket{\ft67 i a, 2iQ} \qquad {\rm and }
\qquad{\cal P}_3\ket{\ft87 i a, 2 iQ}\ ,\eqno(2.48)
$$
where ${\cal P}_3$ is given by (2.45) with
$$
\eqalign{
\tau&={3(p-2i a\over 4i Q},\qquad \mu=-\ft12(p-2i a),\qquad \sigma= - {p-2i
a\over 2i Q}\ ,\cr
\bar\tau&=-{4i a p\over 49},\qquad \bar\mu= {12i a\over 49}, \qquad
\bar\sigma ={-4i a(p-2i a)\over 49}\ ,\cr
\rho&=1,\qquad \bar\rho=-{3p+2i a\over 4i Q},\qquad \kappa={a(p-2i a)\over 2Q},
\qquad \bar\kappa={3\over 2i Q}\ .\cr}\eqno(2.49)
$$

\bigskip
\noindent\undertext{\it Level $4$}
\bigskip

     Again, we consider only the two-scalar $W_3$ string at this level.
There are 20 polarisation-tensor components at level 4, and the
physical-state conditions give rise to 7 solutions.  Three of these are
case-1 states:
$$
{\cal P}_4\ket{\ft{5}7 i a, \ft{15}7 iQ}\ ,\qquad
{\cal P}_4\ket{\ft{9}7 i a, \ft{15}7 iQ} \qquad {\rm and}\qquad
{\cal P}_4\ket{0, 2iQ}\ ;\eqno(2.50)
$$
three are case-2 states:
$$
{\cal P}_4\ket{\ft{20}7 i a,\ft{10}7 i Q}\ ,\qquad
{\cal P}_4\ket{\ft{18}7 i a, \ft{12}7 iQ}
 \qquad {\rm and}\qquad
{\cal P}_4\ket{2 i a, 2 iQ}\ ;\eqno(2.51)
$$
and one is a case-3 (``ordinary'') state:
$$
{\cal P}_4\ket{3 i a,  iQ}\ .\eqno(2.52)
$$
The forms of the polarisation tensors are quite complicated here, and we
shall not give them.

\bigskip\bigskip
\noindent
{\bf 3. Norms}
\bigskip\bigskip

     Having found the solutions to the physical state conditions, one must
still ascertain which of these solutions truly represent physical states and
which are just gauge artifacts, analogous to the longitudinal photons in QED.
The surest way to do this is to consider which states are BRST cohomologically
trivial and which are non-trivial, an issue to which we shall return in section
4.  In ordinary gauge theories, one is accustomed to the fact that the gauge
artifacts, or ``spurious'' states, are also of zero norm, as a result of the
vanishing inner product of their polarisation tensors.  The identification
between spurious states and zero-norm physical states will be seen to persist
in
the case of the $W_3$ string, but the reason for the zero norms of these states
is different from that in ordinary gauge theories or in the critical bosonic
string.  The difference lies in the properties of the Fock space inner product
and in the changed requirements of momentum conservation in the presence of
background charges, and not in the properties of particular polarisation
tensors.

     Let us for simplicity consider at first a theory with only one
species of oscillator $\a_n$ that has a background charge $Q$.  As mentioned
previously the hermiticity condition is $\a_n^\dagger=\a_{-n}$ when $n\ne0$,
and $\a_0^\dagger=\a_0-2iQ$.  The notion of the adjoint corresponding to this
hermiticity condition is the standard one of conformal field theory, in which
the ${\rm SL}(2,\C)$-invariant ket vector $\ket{\Omega}$ is mapped into the
${\rm SL}(2,\C)$-invariant bra vector $\bra{\Omega^\ast}$.  The conditions of
${\rm SL}(2,\C)$ invariance require that $\ket{\Omega}$ have a
{\it right}-eigenvalue of the momentum operator $\a_0$ given by $p_\Omega=0$,
while the {\it left}-eigenvalue of $\a_0$ on $\bra{\Omega^\ast}$ is
$p_{\Omega^\ast}=2iQ$. Given a ket $\ket{p}$ with right-momentum $p$, the
conjugated bra $\bra{p}$ has right-momentum $p^\ast+2iQ$. Note that although
the
adjoint operation maps between vectors of different left- and right-momentum,
it
nonetheless satisfies the condition that for either purely imaginary $p$ or
for $p$ of the form $p={\rm real}+iQ$, the conformal eigenvalue $h=\ft12
p(p-2iQ)$ is the same real value for the conjugated bra and ket vectors.
The eigenvalues $p$ of $\a_0$, and even of the hermitean operator
$\hat\a_0=\a_0-iQ$, need not be real. We must, however, require that the
eigenvalues of $L_0=\ft12\a_0(\a_0-2iQ)+\cdots$ be real. This implies that
$p$ is either purely imaginary or $p=iQ+{\rm real}$.  Thus, the conformal
eigenvalue $h$ is the same for conjugated bras and kets for all physically
allowable values of the momentum.

     From the above, given two momentum states, $\ket{p}$ and $\ket{p'}$, then
we have that
$$
\bra{p'}\a_0\ket{p}=p\braket{p'}{p}=( {p'}^*+2iQ)\braket{p'}{p}\ .
\eqno(3.1)
$$
The momentum conservation law is then that $\braket{p'}{p}$ vanishes
unless
$$
p={p'}^*+2iQ\ .\eqno(3.2)
$$
When the eigenvalues $p$ of $\a_0$ are purely imaginary, the only value of $p$
for which $\braket{p}{p}$ is non-vanishing is $p=iQ$. If we have two states
$\ket{p'}$ and $\ket{p}$ for which $p \ne iQ$ and ${p'}^*=p-2iQ$, then the
inner-product is off-diagonal and so we will have one positive-norm
and one negative-norm state.

     An analogous situation arises in the Liouville theory formulation of the
ordinary string in a non-critical spacetime dimension.  It is well-known
[13,14] that the free-field realisation of Liouville theory actually has
twice as many states as the true nonlinear Liouville theory itself.  This
happens because the presence of the Liouville interaction $e^{g\phi}$ has
the effect of restricting the physical states to only one linear combination
of the $\phi$-momentum eigenstates, where $\phi$ is the Liouville mode,
periodic in the spatial worldsheet coordinate as is appropriate for closed
string theory.  Thus, the true Liouville spectrum is obtained by a
truncation of the spectrum of the free-field realisation.  In the present
case of the single free scalar, the analogous truncation amounts to making a
pairwise identification of the states with  momenta $p$ and $p'$ for which
${p'}^*=p-2iQ$, {\it i.e.}\ one takes only the symmetric linear combination
of the two states of the original free-field theory.  In this truncation,
the half of all the states having negative norm is set to zero, while the
other half with positive norm is retained.

     Although we shall not consider in detail the possibility of introducing
interaction terms analogous to the Liouville interaction $e^{g\phi}$, it is
worth noting that such terms are in principle necessary in order to make inner
product calculations such as (3.1) well-defined.  In the case of the Liouville
theory, this interaction is obtained from a cosmological term $\sim
\int d^2\sigma\sqrt{\gamma}$ for the world-sheet metric $\gamma_{ij}$ upon
gauge
fixing.  The integrand of this term is of conformal weight one, and its
presence dominates the integration over the zero-mode of $\phi$, rendering the
integral convergent.

     For the two-scalar realisation of the $W_3$ algebra, ``screening charge''
terms analogous to the cosmological term of Liouville theory have been found in
[7,9], with integrands of the form $\exp(ge^i_a\phi_i)$,
where $e^i_a$, $a=1,2$ is the $a$'th simple root of $SU(3)$.  A more careful
discussion of the properties of inner products and norms in the $W_3$ string
should properly include such terms, which are presumably responsible for the
truncation of the free-field spectrum that excludes the negative-norm states.

      For $W_3$ strings, we have two background charges, associated with
$\a^{(2)}_n$ and $\a^{\hat\mu}_n$.  We also find that the eigenvalues of
$\a^{(2)}_0$ for physical states are always purely imaginary.  In fact they
are integer multiples of $\ft17 iQ$.  At level $0$, we
find that physical states have momentum $\beta=iQ, \ft67 iQ, \ft87 iQ$.  We
thus identify the states with $\ft67 iQ$ and $\ft87 iQ$, and at the same
time we make a similar identification of $p^{\hat\mu}$ and
$p^{\hat\mu}-2i a^{\hat\mu}$.  At this level we find, therefore, only
two states instead of three.  That we can match up the  states at level $0$
is a consequence of the fact that the zero-mode part of  $L_0$ is symmetric
under $\beta\rightarrow \beta -2iQ$ and the zero-mode part  of $W_0$ changes
by a minus sign under this symmetry.  The sign change is  immaterial since
the $W_0$ intercept is zero.  However, the symmetry under $\beta\rightarrow
\beta-2iQ$ does not extend to the full $W_0$, nor even to  $L_n$ when $n\ne
0$.  Thus one cannot always match up the states at higher levels.  An
exception to this is provided by the ``ordinary'' states of case 3, where
$\beta$ always takes the three values $iQ,\, \ft67 iQ$ and $\ft87 iQ$,
regardless of the level number.  The first is self-conjugate, and we can
match up the latter two states, which are momentum conjugates of each
other, and which have the same $\widetilde L_0$ intercept of $1$.   For
example, at level 1, aside from the ordinary states (case $3$), we find
states (2.30, 2.31) with $\beta=\ft{10}7 iQ,\, \ft{11}7 iQ$ that cannot be
matched up. It follows that these states have zero scalar
product with all other states, since momentum conservation cannot be satisfied
in any two-point function containing them.

     We shall show in the next section that
these zero-norm states are also null, in the sense that they arise from
$L_{-n}$'s and $W_{-n}$'s acting on highest-weight states. At level 2 we find
ordinary states as well as states with $\beta=\ft{10}7 iQ,\, \ft{11}7 iQ $ and
$\ft{12}7 iQ$, and a similar conclusion is obtained.  We also find similar
results at levels 3 and 4, although there are subtleties at level 4 that we
shall discuss at the end of the next section.

\bigskip\bigskip
\noindent{
\bf 4. Null states}
\bigskip\bigskip

     The physical states are by definition those that satisfy the on-shell
conditions given by (2.13).  Included amongst these states are those that have
zero scalar product with all other physical states including themselves; such
states are called null.  Null states arise as a consequence of gauge
invariance.
The remaining non-null physical states describe the true on-shell degrees of
freedom of the string theory.  In a unitary theory these states must have
positive norm.  Some evidence for the unitarity of the physical states for
$W_3$ and $W_N$ strings has been given in ref. [11].   We now wish to find
which of the physical states found in the previous section are null and thus
determine the true degrees of freedom of the $W_3$ string.

    In a general string theory, null states can be written as generators of
the string symmetry algebra acting on highest-weight states.  For the $W_3$
string this means that a null state takes the form
$$
\sum_{\{n_i\}}\sum_{\{m_j\}}L_{-n_1}L_{-n_2}\cdots L_{-n_p}
W_{-m_1}W_{-m_2}\cdots W_{-m_q}\ket\Omega\ ,\eqno(4.1)
$$
where $\ket\Omega$ is a highest-weight state, defined by
$$
\eqalign{
L_0\ket\Omega &=h\ket\Omega, \qquad W_0\ket\Omega=\omega\ket\Omega\ ,\cr
L_n\ket\Omega &=W_n\ket\Omega=0,\qquad n\ge 1\ .\cr}
\eqno(4.2)
$$
The level of the null state is $N^{\rm null}=(\sum_{i=1}^p n_i+\sum_{j=1}^q
m_j)$, and we can study null states level by level.  Note that we distinguish
between the total level number $N$, and $N^{\rm null}$, since although
$\ket\Omega$ is a highest-weight state, it can itself have a level number
$N^{\Omega}$.  Indeed we have $N=N^{\rm null}+N^\Omega$.  At a given level
$N^{\rm null}$, we write down all possible terms and then demand that the
state be physical.  This process, at times tedious, will generate a few
operators which are capable of producing all null states. At level $N^{\rm
null}=1$, we find the following two general null states
$$
\ket{N_1,\pm}=(L_{-1}\pm\ft87 QW_{-1})\ket{\Omega_1,\pm}\equiv
G_1^{\pm}\ket{\Omega_1,\pm}\ ,\eqno(4.3)
$$
with
$h=3$ and $\omega=\mp\ft27 Q$.  At level $N^{\rm null}=2$ we find
$$
\ket{N_2}=\big(L_{-2}+\ft94L_{-1}^2-2W_{-1}^2\big)\ket{\Omega_2}\equiv
G_2\ket{\Omega_2}\ ,\eqno(4.4)
$$
with $h=2$ and $\omega=0$.  There exist two null states at level $N^{\rm
null}=3$, but the corresponding $G_3$ operators can be shown to factorise,
and be expressible, for example, as  $G_1^\pm$ on highest-weight states:
$$
\ket{N_3,\pm}=G_1^\pm(84L_{-2}\pm
39QW_{-2}\mp 24QL_{-1}W_{-1}+ \ft{147}8
L_{-1}^2+21W_{-1}^2)\ket{\Omega_3,\pm}\equiv G_3^\pm \ket{\Omega_3,\pm}\ ,
\eqno(4.5)
$$
where $h=1$ and $\omega=\pm 2Q$.  At level 4, we find three null states,
which can be expressed in terms of factorised operators as follows:
$$
\ket{N_4,\pm}=G_2(48 L_{-2} \pm \ft{100}7
Q W_{-2} +\ft92 L_{-1}^2 \pm \ft{240}{49} a Q L_{-1}W_{-1} + 4W_{-1}^2 )
\ket{\Omega_4,\pm}\equiv G_4^\pm \ket{\Omega_4,\pm}\ ,\eqno(4.6)
$$
with $h=0$, $\omega=\mp \ft{20}7 Q$, and
$$
\eqalign{
\ket{N_4,0}&=G_1^-(12 L_{-3} -12 L_{-1}L_{-2} -\ft{147}8
L_{-1}^3 -3 W_{-1} W_{-2} +\ft{24}7 Q W_{-1}^3\cr
&\qquad -63Q L_{-1} W_{-2} -21 Q
L_{-1}^2 W_{-1} +\ft{96}7 Q L_{-2} W_{-1} + 3L_{-1} W_{-1}^2)
\ket{\Omega_4,0}\cr
&\equiv G_4^0 \ket{\Omega_4,0}\ ,\cr}\eqno(4.7)
$$
with $h=0$ and $\omega=0$.  Thus all of these kinds of states, being null,
need not be considered in the physical spectrum.

     It is interesting to
compare the null states of the usual Virasoro string with the ones of the
$W_3$ string. In the case of the one-scalar Virasoro string, the null states
occur only at levels $N^{\rm null}=1, 2, 5, 7,12, 15, \ldots $. There are
only two fundamental null operators, namely $L_{-1}$ and $\tilde
L_{-2}=L_2+\ft32L_1^2$; all higher-level null operators can be composed as
one or other of these two operators acting on highest-weight states.  In the
case of the two-scalar $W_3$ string, there exist null states at any level
$N^{\rm null}$.  To see this, note that the number of currents in the $W_3$
string is 2, namely the spin-2 current $T$ and the spin-3 current $W$. Thus
all the physical states of the two-scalar $W_3$ string can be expressed, by
a transformation of basis, as states of the form given in (4.1). (We assume
for now that this transformation of basis is non-singular.  See later,
however.)  Indeed all the excited physical states of the two-scalar $W_3$
string are null, just as all the excited physical states of the one-scalar
Virasoro string are null.

     To solve for the actual null states explicitly, we must solve the
conditions on the highest-weight states such as $\ket{\Omega_1,\pm}$,
$\ket{\Omega_2}$, $\ket{\Omega_3,\pm}$, $\ket{\Omega_4,\pm}$ and
$\ket{\Omega_4,0}$.  It can happen that at a given level there exists no
solution; however, if a solution does exist it provides a null state for the
 $W_3$ string.   We shall find the actual null states level by level using
the processes described above.  At level $N=0$, clearly there are no null
states, so all the physical states represent degrees of freedom of the
system.  Of course taking into account the identification of states with
conjugate momenta, there are two, rather than three, degrees of freedom.  At
level $N=1$, the null states can only arise from  $\ket{N_1,\pm}$ acting on
$\ket{\Omega_1,\pm}$ which has no excitations.   Thus $\ket{\Omega_1,\pm}$
is of the form $\ket{\beta, p^{\hat\mu}, \pm}$.   Using the expression for
$W_0$ in equation (2.14), with the appropriate $h$  and $\omega$ given below
equation (4.3), we find that $\ket{\Omega_1,+}$ can  only have the $\beta$
values
$$
\beta=\ft{11}7 iQ,\, \ft67 iQ,\, \ft47 iQ\ ,\eqno(4.8)
$$
while $\ket{\Omega_1,-}$ can only have $\beta$ equal to
$$
\beta=\ft{10}7 iQ,\, \ft37 iQ,\, \ft87 iQ\ .\eqno(4.9)
$$
The actual null states are then given by
$$
\eqalign{
\ket{N_1,+}\propto \left\{\a^{(2)}_{-1}\Big (6-2\beta(\beta-\ft{10}7 iQ)
\Big) +2 p_{\hat\mu}(\beta-\ft47 iQ)\a^{\hat\mu}_{-1}\right\}
\ket{p^{\hat\mu},\beta}\ ,\cr
\ket{N_1,-}\propto \left\{\a^{(2)}_{-1}\Big (6-2\beta(\beta-\ft{11}7 iQ)
\Big) +2 p_{\hat\mu}(\beta-\ft37 iQ)\a^{\hat\mu}_{-1}\right\}
\ket{p^{\hat\mu},\beta}\ .\cr}\eqno(4.10)
$$
It is easy to check that the states for which $\beta=\ft47 iQ$ and $\beta=
\ft37 iQ$ vanish.  Substituting the remaining allowed values of $\beta$ of
equations (4.8) and (4.9) into equation (4.10), we recover all the physical
states of the two-scalar $W_3$ string given in equations (2.35) and (2.36).
This explicitly verifies that all the states of the two-scalar $W_3$ string
are null at this level.   For the $(D+2)$-scalar $W_3$ string we find that
part of the states given in equation (2.33) with $\beta=\ft67 iQ,\,\ft87 iQ$
are null and so they are subject to the gauge invariance $\xi^{\hat\mu}
\rightarrow \xi^{\hat\mu}+\Lambda p^{\hat\mu}$,  This is analogous to the
Virasoro string, since in this branch the effective intercept $a^{\rm eff}$
is equal to $1$; gauge invariance at this level implies the presence of
massless states.   All the states of equation (2.31) with $\beta=\ft{10}7
iQ,\,\ft{11}7 iQ$ are null and therefore can be gauged away.  Thus if
we take into account the identification of the states with $\beta=\ft67
iQ$ and $\ft87 iQ$, and the constraints given by equation (2.34) and the
above-mentioned gauge invariance, there exist $(D+1)-1 +(D+1)-2=2D-1$
degrees of freedom at level one.

     At level two, null states can arise either from $G_1^{\pm}$ acting on
level-1 highest-weight states or from $G_2$ acting on level-0 states, with
conformal weight and $W$ weight given below equations (4.3) and (4.4)
respectively.  Let us consider the latter first;  applying $W_0$ of
equation (2.14), with $h=2$ and $\omega=0$, we find that the only allowed
values of $\beta$ are
$$
\beta=\ft27 iQ,\, iQ,\, \ft{12}7 iQ\ .\eqno(4.11)
$$
Evaluating $\ket{N_2}$ of equation (4.4), we find that it vanishes if
$\beta=\ft27 iQ$. For $\beta=iQ$, it gives
$$
\ket{N_2}\propto \left\{\a_{-2}^{\hat\mu}(7p_{\hat\mu}+3i a_{\hat\mu})
+\a_{-1}^{\hat\mu}\a_{-1}^{\hat\nu}(4p_{\hat\mu}p_{\hat\nu}+\ft32
\eta_{\hat\mu\hat\nu})\right\} \ket{p^{\hat\mu},iQ}\ .\eqno(4.12)
$$
For $\beta=\ft{12}7 iQ$ the null states are
$$
\eqalign{
\ket{N_2}\propto\Big( &175\a^{(2)}_{-1}\a^{(2)}_{-1}+\ft{1400}7 iQ
p_{\hat\mu}\a^{(2)}_{-1}\a^{\hat\mu}_{-1}-\ft{296}7 iQ\a^{(2)}_{-2}
\cr
&-70(p_{\hat\mu} -i a_{\hat\mu}) \alpha^{\hat \mu}_{-2}+
\a^{\hat\mu}_{-1}\a^{\hat\nu}_{-1}(140p_{\hat\mu}p_{\hat\nu} -
33\eta_{\hat\mu  \hat\nu})\Big)\ket{p^{\hat\mu}, \ft{12}7
iQ}\ .\cr}\eqno(4.13)
$$
The states of equation (4.13) can be used to gauge
away the $\beta=\ft{12}7  iQ$ state of equation (2.41). The null states
given in (4.12) imply  the gauge invariance
$$
\eqalign{
\xi_{\hat\mu}&\rightarrow \xi_{\hat\mu}+(7p_{\hat\mu}+3i a_{\hat\mu})\Lambda
\ ,\cr
\varepsilon_{\hat\mu\hat\nu}&\rightarrow \varepsilon_{\hat\mu\hat\mu}+
(4p_{\hat\mu}p_{\hat\nu}+\ft32\eta_{\hat\mu\hat\nu})\Lambda\cr}
\eqno(4.14)
$$
of the ``ordinary'' states of equation (2.38).  Thus talking into account
the constraints of equation (2.40) and this gauge invariance, the states with
$\beta=iQ$ have $(D+1)(D+2)-2$ degrees of freedom.

      Secondly, we look at the null states at level 2 arising from
null-state operators $G_1^{\pm}$ acting on the level-1 highest-weight states
$\ket{\Omega_1,\pm}$, which take the form
$$
\ket{\Omega_1,\pm}=\xi_{\hat\mu}\a^{\hat\mu}_{-1}\ket{p^{\hat\mu},\beta}
\eqno(4.15)
$$
provided that
$$
(p_{\hat\mu}-2i a_{\hat\mu}) \xi^{\hat\mu}=0\ .\eqno(4.16)
$$
For the choice of ``$+$'' in (4.15), we have $\beta=\ft47 iQ, \ft67 iQ$ and
$\ft{11}7 iQ$; For the ``$-$'' choice, we have $\beta= \ft37 iQ, \ft87 iQ$
and $\ft{10}7 iQ$.  The null states $\ket{N_1,\pm}$ are then found to be of
the form
$$
\eqalign{
\ket{N_1,\pm}\propto\Big[ &\Big(
\beta(\beta-i Q)-p^{\hat\mu}(p_{\hat\mu}-2i a_{\hat\mu})-2\pm\ft17 iQ\Big)
\a^{(2)}_{-1}\xi_{\hat\mu}\a^{\hat\mu}_{-1}\cr
&+(-2\beta+iQ\pm\ft17 iQ)(p_{\hat\mu}\xi_{\hat\nu}\a^{\hat\mu}_{-1}
\a^{\hat\nu}_{-1}+\xi_{\hat\mu}\a^{\hat\mu}_{-2})\Big]
\ket{p^{\hat\mu},\beta}\ .}\eqno(4.17)
$$
For the ``$+$'' sign we find that the states with $\beta=\ft47 iQ$ vanish
identically, while $\beta=\ft67 iQ$ and $\beta=\ft{11}7 iQ$ lead
respectively to the states
$$
\eqalignno{
\ket{N_1,+}&\propto (p_{\hat\mu}\xi_{\hat\nu}\a^{\hat\mu}_{-1}
\a^{\hat\nu}_{-1} +\xi_{\hat\mu}\a^{\hat\mu}_{-2})\ket{p^{\hat\mu},
\ft67 iQ}\ ,&(4.18a)\cr
\ket{N_1,+}&\propto\Big(13\xi_{\hat\mu}\a^{(2)}_{-1}\a^{\hat\mu}_{-1} +
2iQ(p_{\hat\mu}\xi_{\hat\nu}\a^{\hat\mu}_{-1}\a^{\hat\nu}_{-1} +
\xi_{\hat\mu}\a^{\hat\mu}_{-2})\Big)\ket{p^{\hat\mu},\ft{11}7 iQ}\ .
&(4.18b)\cr}
$$
Similarly, for the ``$-$'' sign we find that that states given in (4.17)
vanish if $\beta=\ft37 iQ$, while they are non-zero for $\beta=\ft87 iQ$
and  $\beta=\ft{10}7 iQ$, which leads to the states
$$
\eqalignno{
\ket{N_1,-}&\propto \Big(p_{\hat\mu}\a^{\hat\mu}_{-1} \xi_{\hat\nu}
\a^{\hat\nu}_{-1}+\xi_{\hat\mu}\a^{\hat\mu}_{-2}\Big)\ket{p^{\hat\mu}, \ft87
iQ}\ ,&(4.19a)\cr
\ket{N_1,-}&\propto \Big( 7\xi_{\hat\mu}\a^{(2)}_{-1}\a^{\hat\mu}_{-1}
+4iQ\xi_{\hat\nu}p_{\hat\mu}\a^{\hat\mu}_{-1} \a^{\hat\nu}_{-1} +
\xi_{\hat\mu}\a^{\hat\mu}_{-2}\Big) \ket{p^{\hat\mu},\ft{10}7 iQ}\ .
&(4.19b)\cr}
$$
The null states of equation (4.18$b$) and (4.19$b$) can be used to gauge
away the physical states of equation (2.44).  It follows from (2.40) that
the  null states of equation (4.18$a$) and (4.19$a$) imply the gauge
invariance
$$
\xi_{\hat\mu}\rightarrow\xi_{\hat\mu} + \Lambda_{\hat\mu}\qquad
\varepsilon_{\hat\mu\hat\nu}\rightarrow \varepsilon_{\hat\mu\hat\nu} +
\ft12(p_{\hat\mu}\Lambda_{\hat\nu} +
p_{\hat\nu}\Lambda_{\hat\mu})\eqno(4.20)
$$
of the physical states given in equation (2.39).  Thus these states have
$(D+1)(D+2)/2-D-1$ degrees of freedom, taking into account the
identification, and the constraint equation (2.40), and that
$(p-2i a)_{\hat\mu}\Lambda^{\hat\mu}=0$.

      We note that all the degrees of freedom originate from states that are
of ordinary (case 3) type, and that they give $(D+1)(D+2)-D-3$ degrees of
freedom in all at level 2.  For the two-scalar $W_3$ string no null state
arises from $G_1^{\pm}$ acting on a level-1 highest-weight state
$\ket{\Omega_1,\pm}$. However, one can obtain null states by acting with
$G_2$ on the primary states $\ket{p,\beta}$ with $\beta=\ft27 iQ, iQ,
\ft{12}7 iQ$.  When $\beta=\ft27 iQ$, $G_2$ annihilates $\ket{p,\beta}$,
while for other two %\beta$ values, it will give the states given below
(2.44).   Thus this verifies that all the states of two-scalar $W_3$ string
at level 2 are null.

   At levels 3 and 4, we only consider the two-scalar realisation.  At
level 3, it can be shown that all the physical states, namely of equation
(2.46) and (2.48), can be  rewritten as $G_3$ acting on primary states
$\ket{p, \beta}$ with  $\beta=\ft{11}7 iQ, \ft{12}7 iQ, 2iQ$.  In other
words, they can be written  as $G_1^+$ acting on level-2 highest-weight
states with $h=3$ and  $\omega=-\ft27 Q$.  Thus the physical states at
level 3 are all null for the two-scalar $W_3$ string.  At level 4, a new
feature emerges.  The first and second states in (2.50) can be written in
terms of $G_4^-$ in (4.6) acting on $h=0$, $\omega=\ft{20}7 Q$
highest-weight states.  The first and second states in (2.51) can be
written in terms of $G_4^+$ in (4.6) acting on $h=0$, $\omega=-\ft{20}7 Q$
highest-weight states.  The third state in (2.51) can be written in terms
of $G_4^0$ in (4.7) acting on an $h=0$, $\omega=0$ highest-weight state.
However, the remaining two level-4 states, namely the third state in (2.50)
and the state in (2.52), cannot be written in terms of any $G$-type
operator acting on tachyonic states.  What has happened is that the tacit
assumption that we made earlier in this section, namely that the
transformation of basis from $\del\varphi_1$, $\del\varphi_2$ to $L_{-n}$,
$W_{-n}$ is non-singular, has broken down in these cases.  It is still the
case that these two states are null however; we have explicitly checked
that they can be written as BRST variations of some particular states.
Unlike the ordinary string, where any BRST-trivial state can be written in
terms of $L_{-n}$ descendants of some highest-weight state, for the $W_3$
string one can no longer write all BRST-trivial states in terms of $L_{-n}$
and $W_{-n}$ descendants of highest-weight states.

\bigskip\bigskip
\noindent{
\bf 5. A toy model; the one-scalar string }
\bigskip\bigskip

     In this section we consider a bosonic string consisting of one
coordinate $\phi$ in the presence of a background charge.  The
energy-momentum tensor is given by $T=-\ft12 (\partial \phi)^2
-Q\partial^2\phi$.  To cancel the conformal anomaly, we require that
$c=26=1+12 Q^2$, {\it i.e.}\ $Q^2=\ft{25}{12}$.  The physical states are
given by the usual on-shell conditions
$$
(L_0-1)\ket\psi=0\ ,\qquad {\rm and}\qquad L_n\ket\psi=0 \qquad
{\rm with}\qquad n \ge 1\ .\eqno(5.1)
$$

      We wish to find all the physical states of this one-dimensional
string, and also to construct the states in terms of screening operators acting
on the vacuum.  This exercise proves to be a very useful model for the
two-scalar $W_3$ string that will be considered in section 5.   It is
straightforward to analyse the lowest levels of the one-dimensional bosonic
string.  At level $0$ we find two physical states,
$$
\ket{\ft65 iQ}\qquad {\rm and}\qquad \ket{\ft45 iQ}\ ,\eqno(5.2)
$$
where $\ket{p}$ denotes a momentum eigenstate satisfying $\a_n\ket{p}=0$ for
$n\ge 1$, and $\a_0\ket{p}=p\ket{p}$.  At level $1$, we have only one
physical state
$$
\a_{-1}\ket{2iQ}\ .\eqno(5.3)
$$
At level $2$ there is also only one physical state
$$
(\ft65 iQ\a_{-1}^2+\a_{-2})\ket{\ft{12}5 iQ}\ .\eqno(5.4)
$$
To find the physical states of the string at higher levels, it is useful to
write a general state in the form
$$
L_{-n_1}L_{-n_2}\cdots L_{-n_p}\ket{\beta}\ .\eqno(5.5)
$$
Note that we have replaced $\a_{-n}$ by $L_{-n}$ in generating states.  For
a generic state one can always make this change, since for a single-scalar
realisation the $\alpha_{-n}$'s and $L_{-n}$'s are generically in
one-to-one  correspondence.  However, for some very special momentum $\beta$
it may happen that the $L_{-n}$'s span only a subspace.  In this context we
note, for example, that $L_{-n}\ket{\beta=0}$ contains no term linear in the
oscillator $\a_{-n}$.  For this particular section, however, we shall
tacitly assume that for the on-shell momentum this possibility does not
arise.  Clearly all physical states, except the two tachyons of equation
(5.2), will be null.

    We recall [15,16] that a highest-weight state $\ket{\chi}$ of weight
$\Delta$ will itself have a highest-weight state in its Verma module if
$$
\Delta={c-1\over 24} + \ft14 (n\a_+ + m\a_-)^2\ ,\qquad n,m \in Z_+\ ,
\eqno(5.6)
$$
where $c$ is the central charge and
$$
\a_{\pm}={\sqrt{1-c}\pm\sqrt{25-c} \over \sqrt{24}}\ .\eqno(5.7)
$$
The highest-weight state occurs at level $N=n m$.   Now for the bosonic
string with $c=26$, a null state has the form
$$
L_{-n_1}L_{-n_2}\cdots L_{-n_p}\ket{\Omega_N}\ ,\eqno(5.8)
$$
where $\ket{\Omega_N}$ satisfies $L_n\ket{\Omega_N}=0$ for $n\ge 1$, and
$(L_0+N-1)\ket{\Omega_N}=0$ with $N=\sum_i n_i$.  Such a states will be
physical if there exist positive integers such that
$$
\Delta_{n,m}+n m=1\ , \qquad {\rm {\it i.e.\ }}\qquad (3n-2m)^2=1\ ,
\eqno(5.9)
$$
It is straightforward to show that the bosonic string will have such
physical null states at levels
$$
\ft12 n(3n\pm 1)\ ,\qquad n=1,2,5,\ldots\ .\eqno(5.10)
$$
The actual null states of the theory occur whenever the state $\ket{\Omega}$
admits a solution.  Clearly, we can take $\ket{\Omega_N}=\ket\beta$, with
$\beta(\beta-2iQ)=2(1-N)$, to construct null states at the levels given in
(5.10).  However we shall also find other null states.

     The above discussion holds for any bosonic string, since we used the
abstract generators $L_{-n}$ to construct null physical states.  For the
one-scalar bosonic string we can similarly analyse when highest-weight
states $\ket{\Omega_N}$ exist.  For example, let us consider
$\ket{\Omega_1}$, which can be written as
$L_{-m_1}L_{-m_2}\cdots\L_{-m_q}\,\ket{\Gamma_{1,M}}$ where $M=\sum_i m_i$,
if there exist $n,m\in Z_+$ such that $\Delta_{n,m}+nm=0$, {\it i.e.\ }
$(3n-2m)^2=25$.  Analysing this result, we find that it leads to null
physical states in the original bosonic string at levels
$$
\ft12 n(3n\pm 1)\ , \qquad n=2,4,6,\ldots\ .\eqno(5.11)
$$
In fact, there are solutions of $\ket{\Omega_N}$ for arbitrary $N$, but they
lead to null states at the same levels as those given by $\ket{\Omega_1}$.
In principle one can look for null states within null states, but one finds
no additional states.  Putting the results of equations (5.10) and (5.11)
together, we find that the one-scalar string has null states at levels
$$
\ft12 n(3n\pm 1)\ ,\qquad n \in Z_+\ .\eqno(5.12)
$$
It follows from the usual on-shell condition that the momentum $\beta$ is
quantised for all physical states since the $\beta$ values are $\ft15 iQ$
multiplied by certain integers, {\it i.e.\ }
$$
\beta=\Big(5\pm (6n\pm 1)\Big) \ft15 iQ\ .\eqno(5.13)
$$
This can be better understood by a screening-charge analysis that we shall
discuss presently.

     It is quite instructive to examine in detail the way in which the null
physical states of the one-scalar string can be constructed by acting with
$L_{-n}$'s on a highest-weight state.  As is well known, if $\ket{h}$ is a
highest-weight state with weight $h=0$, {\it i.e.}\ $L_0\ket{h}=0$, then it
gives a highest-weight null state $L_{-1}\ket{h}$ that has weight 1.  This
is precisely a level-1 physical null state of string theory. In fact only if
$h=0$ does one get a highest-weight state by this means.  At level 2, one
finds that if $c=26$ and $\ket{h}$ is a highest-weight state then the
operator
$$
\tilde L_{-2}=L_{-2}+\lambda L_{-1}^2 \eqno(5.14)
$$
will give a null highest-weight state $\tilde L_{-2}\ket{h}$ provided
that either $h=-1$, in which case $\lambda=\ft32$, or $h=-\ft{13}8$, in
which case $\lambda=\ft23$.  The former case is relevant to string theory,
since then the null state $\tilde L_{-2}\ket{h}$ will be highest weight with
weight $-1+2=1$, thus satisfying the physical state conditions.  In
Appendix A we present results for the first twelve levels in string theory,
{\it i.e.}\ we give the results for the operators $\tilde L_{-n}$ for
$n=1,2,\ldots,12$.

      We now use the screening charges to construct explicitly the physical
states of the one-dimensional string.  The two screening charges are
$$
S_\pm=\oint e^{i\a_\pm \phi (z)}\ ,\eqno(5.15)
$$
where $\a_+=\ft65 iQ$, $\a_-=\ft45 iQ$.  The two level-$0$ states of
equation (4.2) are $e^{i\a_\pm\phi (0)}\ket{0}$. Acting with $S_\pm$ on a
physical state of momentum $p$ will be well defined if $p\cdot\a_\pm$ is a
negative-definite integer; such a state will also be physical.
We find in this way the physical states
$$
(S_+)^n\ket{\a_+}\ ,\qquad n\in  Z_+\ ,\eqno(5.16)
$$
which have momentum $(n+1)\a_+$ and occur at levels $\ft12 n(3n+1)$. We also
find physical states at levels $\ft12 n(3n-1)$, namely
$$
(S_+)^n\ket{\a_-}\ ,\qquad n\in  Z_+\ ,\eqno(5.17)
$$
which have momenta $n\a_++\a_-$.  The only other well-defined states come
from $S_-$ acting on $(S_+)^n\ket{\a_+}$, which, however, gives rise to the
same physical states as those of equation (5.16).  We note that we get
precise  agreement with the levels at which null states occur as predicted
by the Kac formula, {\it i.e.}\ those of equation (5.12).  This is perhaps
not too surprising  since the highest-weight states of the Verma module can
be constructed from  screening operators.  The screening-charge analysis also
gives us more  understanding as to why the physical states of the
one-dimensional string are null.  As discussed in section 3, a physical
state will have zero scalar product with all other physical states if there
does not exist its ``dual'' partner in the physical spectrum. Since the
momenta of the screening operators are positive, it follows that the momenta
of the physical states will increase monotonically as the level number
increases.  Therefore no higher-level physical states can occur in the
momentum-conjugate pairs necessary for a non-vanishing scalar product.

\bigskip\bigskip
\noindent{
\bf 6. Screening operators and the two-scalar $W_3$ string }
\bigskip\bigskip

     In section 2, we found all the physical states up to level $4$ in the
two-scalar $W_3$ string.  In section 3, we showed that all these states
above  level zero are null.  Here we shall construct these states in terms
of  screening operators acting on the vacuum.

     The $W_3$ algebra possesses four screening charges [7]
$$
S_{i\pm}=\oint e^{i\a_\pm {\bf e}_i\cdot \vec\phi(z)}\ ,\qquad\qquad
i=1,2\ ,\eqno(6.1)
$$
where ${\bf e}_i$ are the two simple roots of $su(3)$, {\it i.e.\ }
${\bf e}_1=(\sqrt{2},0)$, ${\bf e}_2=(-\sqrt{\ft12},\sqrt{\ft32})$, and
$\a_\pm$ satisfy the equation $\a_\pm^2- i\a_0\a_\pm-1=0$.  Here $\a_0$ is a
background charge parameter satisfying $c=2+24\a^2_0$.  For the $W_3$ string
$c=100$, and one finds $\a_+=4i/\sqrt{12}$ and $\a_-=3i/\sqrt{12}$.
Consequently,
$$
\eqalign{
\a_+{\bf e}_1&=(\ft87 i a,\, 0)\ ,\qquad \qquad\a_-{\bf e}_1=(\ft67 i a,\,
0)\ ,\cr \a_+{\bf e}_2&=(-\ft47 i a,\ft47 iQ)\ ,\qquad
\a_-{\bf e}_2=(-\ft37 i a,\ft37 iQ)\ .\cr}\eqno(6.2)
$$

     The screening operators of equation (6.1) commute with $L_n$ and $W_n$, as
is most simply shown by utilising the quantum Miura construction of these
operators based on $su(3)$ [].  A more lengthy calculation shows that these
are the unique screening operators.  Unlike the one-dimensional Virasoro
string, where the screening operators are the same as the tachyon operators,
the six level-$0$ states of the two-scalar $W_3$ string, given by equation
(2.36), can be built from products of four screening operators acting on the
vacuum, provided that one takes an appropriate limit [9].  There are a number
of ways to achieve this result.  One set of choices is
$$
\eqalign{
\vec\gamma_1&=\a_+ {\bf e}_2 + \a_-({\bf e}_2+ 2{\bf e}_1)=
(\ft57 i a, iQ)\ ,\qquad
\vec\gamma_2=\a_+(2{\bf e}_1+{\bf e}_2) + \a_1 {\bf e}_2 =
(\ft97 i a, iQ)\ ,\cr
\vec\gamma_3&=\a_-(2{\bf e}_2+2{\bf e}_1)=
(\ft67 ia, \ft67 iQ)\ ,\qquad\qquad
\vec\gamma_4=\a_+ {\bf e}_1 + \a_-({\bf e}_1+2{\bf e}_2)=
(\ft87 i a, \ft67 iQ)\ ,\cr
\vec\gamma_5&=\a_+ ({\bf e}_1+ 2{\bf e}_2) + \a_- {\bf e}_1 =
(\ft67 i a, \ft87 iQ)\ ,\qquad
\vec\gamma_6=\a_+ (2{\bf e}_1 +2 {\bf e}_2)=(\ft 87 i a, \ft87 iQ)\ .\cr}
\eqno(6.3)
$$
For $\vec\gamma_2$, for example, we construct the corresponding state
$$
e^{i\a_- {\bf e}_2\cdot \vec\phi(z_1)}
e^{i\a_+ {\bf e}_2\cdot \vec\phi(z_2)}
e^{i\a_+ {\bf e}_1\cdot \vec\phi(z_3)}
e^{i\a_+ {\bf e}_1\cdot \vec\phi(z_4)}\ , \eqno(6.4)
$$
set $z_{12}=z_{23}=z_{34}=\epsilon$, where $z_{ij}\equiv z_i-z_j$, and send
$\epsilon\rightarrow 0$.

        We now consider the action of one screening operator on the
level-$0$ states.  The operator $S_{i\pm}$ acting on a state of momentum
${\bf k}$ is well defined if $\a_\pm{\bf k}\cdot{\bf e}^i$ is an integer.
Such a state will be physical if the original state is physical.  Acting
with $S_{i\pm}$ on the six tachyon states, we find four of the six
level-$1$ states;
$$
\eqalign{
S_{1+}\ket{\vec\gamma_3}&=\a_+ {\bf e}_1\cdot\vec\a_{-1}
\ket{ 2i a, \ft67 iQ}\ , \qquad S_{2+}\ket{\vec\gamma_3}=
\a_+{\bf e}_2 \cdot \vec\a_{-1}\ket{\ft27 i  a, \ft{10}7 iQ}\ ,\cr
S_{1+}\ket{\vec\gamma_5}&=\a_+{\bf e}_1\cdot\vec\a_{-1}
\ket{2i a, \ft87 iQ}\ ,\qquad S_{2-}\ket{\vec\gamma_6}=\a_-{\bf e}_2
\cdot\vec\a_{-1}\ket{\ft57 i a, \ft{11}7 iQ}\ ,\cr}
\eqno(6.5)
$$
and two of the three level-$2$ states;
$$
S_{1+}\ket{\vec\gamma_2}=\oint {1\over z^3} e^{i \a_+ {\bf e}_1 \cdot\vec
\phi(z)}\ket{\ft{17}7 i a, iQ}\ ,\qquad
S_{2+}\ket{\vec\gamma_5}=\oint {1\over z^3} e^{i \a_+ {\bf e}_2 \cdot\vec
\phi(z)}\ket{\ft27 i a, \ft{12}7 iQ}\ .\eqno(6.6)
$$

     To get the additional states at levels one and two, we must act with
two screening charges on the tachyon states.  In this case, we shall examine
when the action of two such operators is well defined.  We consider the
quantity
$$
\oint dz_3\oint dz_2 :e^{i{\bf k}_3\cdot\vec\phi(z_3)}: :e^{i {\bf k}_2
\cdot\vec\phi(z_2)}:\ket{\bf p}\ .\eqno(6.7)
$$
Using the usual expression for the normal-ordered exponential, the above
quantity becomes
$$
\oint_{c_3} dz_3\oint_{c_2} dz_2\, (z_3-z_2)^{\bf k_3\cdot k_2} z_3^{\bf
k_1\cdot p} z_2^{\bf k_2\cdot p} :e^{i {\bf k}_3\cdot\vec\phi(z_3) +
i{\bf k}_2\cdot \phi(z_2)}: \ket{\bf p}\ .\eqno(6.8)
$$
Since the $z_3$ exponential comes first, we must have the contour with
$|z_3|>|z_2|$.  We choose the contours so that the $c_2$
contour is always within the $c_3$ contour except at one point where the two
contours touch.  We now substitute $z_2=w z_3$ and carry out the $z_3$
integral first.  We note that $|w|<1$ except when the contours
touch, in which case it is equal to 1.  Equation (5.8) then becomes
$$
\oint dw \oint dz_3\, z_3^{{\bf k}_3\cdot{\bf k}_2 + ({\bf k}_3 +
{\bf k}_2)\cdot{\bf p} + 1} (1-w)^{{\bf k}_3\cdot{\bf k}_2} w^{{\bf
k}_1\cdot{\bf p}} :e^{i{\bf k}_3\cdot\vec\phi(z_3)+ i{\bf k}_2\cdot
\vec\phi(w z_3)}\ket{\bf p}\ .\eqno(6.9)
$$
This integral is well defined if ${\bf k}_3\cdot{\bf k}_2 + {\bf k}_3
\cdot{\bf p} + {\bf k}_2\cdot{\bf p}$ is an integer.  In this case the
$z_3$  integral becomes analytic.  For a general case of the action of $n$
exponentials on a state of momentum ${\bf p}$, a similar argument implies
that the integral is well defined if $\sum_{i>j}^n{\bf k}_i\cdot{\bf k}_j
+\sum_{j=1}^n{\bf k}_j\cdot{\bf p}$ is an integer.

      Returning to the states of equation (5.9), and assuming that ${\bf k}_3
\cdot{\bf k}_2 + {\bf k}_3\cdot{\bf p} + {\bf k}_2\cdot{\bf p}$ is indeed an
integer, we can carry out the $w$ integration, which has the generic form
$$
I(a,b)=\oint dw\, w^a (1-w)^b \ .\eqno(6.10)
$$
The contour of integration encloses the branch cut between
$0$ and $1$ that results from $a$ and $b$ being non-integer.  We can deform
the contour to lie infinitesimally above and below the cut while remaining
on the given Riemann sheet.  However, the section below the cut is
obtained by a $z \rightarrow e^{i2\pi a} z$ rotation and we find that
$$
\eqalign{
I(a,b)&=(e^{2\pi ia}-1)\int^1_0dw\, w^a(1-w)^b\cr
&=(e^{2\pi ia}-1){\Gamma(a+1)\Gamma(b+1)\over\Gamma(a+b+2)}\ .\cr}\eqno(6.11)
$$
We calculate the integral for $a>-1, b>-1$, and then use analytic
continuation to define it for all $a$ and $b$.  We note that the phase
$e^{2\pi ia}$ is the same for all $a$'s that differ by an integer.  Using
the above method, we find that
$$
\eqalign{
S_{1+}S_{2+}\ket{\vec\gamma_4}&=\oint dz_3\oint dz_2\,
z_3^{-8/3}z_2^{-5/3}
(z_3-z_2)^{4/3} \ket{\vec\gamma_4+\a_+({\bf e}_1+{\bf e}_2)}\cr
&=\oint dw\, w^{-8/3}(1-w)^{4/3} i(\a_+ w {\bf e}_1\cdot\vec\a_{-1}
+\a_+{\bf e}_2\cdot\vec\a_{-1})\ket{\vec\gamma_4+
\a_+({\bf e}_1+{\bf e}_2)}\cr
&\propto \a_+(2{\bf e}_2-5{\bf e}_1)\cdot\vec\a_{-1}\ket{\ft{12}7 i a,
\ft{10}7 iQ}\ ,\cr}\eqno(6.12)
$$
which is one of the remaining level-$1$ states.  The other remaining
level-$1$ state is constructed as
$$
S_{1-}S_{2-}\ket{\vec\gamma_5}\propto\a_-(5{\bf e}_2 -2{\bf e}_1 )\cdot
\vec\a_{-1}\ket{\ft97 i a,\ft{11}7 iQ}\ .\eqno(6.13)
$$
The remaining level-$2$ state with momentum $(\ft{12}7 i a, \ft{12}7 iQ)$
is given by
$$
S_{1+}S_{2+}\ket{\vec\gamma_6}\ .\eqno(6.14)
$$

       At level three we find that all the four states can be
constructed as a screening operator acting on the level-one states given in
(6.12) and (6.13), or they can also be viewed as three screening operators
acting on the corresponding tachyonic states.   Those of equation (2.49)
occur as
$$
\eqalign{
S_{2+}{\cal P}_1\ket{\ft{12}7 i a, \ft{10}7 iQ}&=
S_{2+}S_{1+}S_{2+}\ket{\vec\gamma_4}={\cal P}_3\ket{\ft87 i a, 2iQ}\ ,\cr
S_{2-}{\cal P}_1\ket{\ft97 i a, \ft{11}7 iQ}&=
S_{2-}S_{1-}S_{2-}\ket{\vec\gamma_5}={\cal P}_3\ket{\ft67 i a, 2iQ}\ .\cr}
\eqno(6.15)
$$
Those of equation (2.47) are
$$
\eqalign{
S_{1+}{\cal P}_1\ket{\ft97 i a, \ft{11}7 iQ}&=
S_{1+}S_{1-}S_{2-}\ket{\vec\gamma_5}={\cal P}_3\ket{\ft{17}7 i a,
\ft{11}7 iQ}\ ,\cr
S_{1-}{\cal P}_1\ket{\ft{12}7 i a, \ft{10}7 iQ}&=
S_{1-}S_{1+}S_{2+}\ket{\gamma_4}={\cal P}_3\ket{\ft{18}7 i a,\ft{10}7 iQ}
\ .\cr}\eqno(6.16)
$$
The reader may verify that the polarisations in the generic symbol ${\cal
P}_3$ are the same as given in equation (2.47) and (2.49).  In fact, there
are a number of ways to construct the physical states by using different
screening operators acting on different states.

     At level four, we can obtain the seven states listed in (2.50)-(2.52)
as follows:
$$
\eqalign{
S_{2+}S_{2-}S_{1-}\ket{\vec\gamma_5}&=
{\cal P}_4\ket{\ft{5}7 i a,\ft{15}7iQ}\ ,\cr
S_{2+}S_{2+}S_{1+}\ket{\vec\gamma_2}&=
{\cal P}_4\ket{\ft{9}7 i a,\ft{15}7iQ}\ ,\cr
S_{2-}S_{2-}\ket{\vec\gamma_5}&=
{\cal P}_4\ket{0,2iQ}\ ,\cr
S_{2+}S_{1+}S_{1+}\ket{\vec\gamma_4}&=
{\cal P}_4\ket{\ft{20}7 i a,\ft{10}7iQ}\ ,\cr
S_{2+}S_{1+}S_{1-}\ket{\vec\gamma_6}&=
{\cal P}_4\ket{\ft{18}7 i a,\ft{12}7iQ}\ ,\cr
S_{2-}S_{2-}S_{1-}S_{1-}\ket{\vec\gamma_6}&=
{\cal P}_4\ket{2i a,2iQ}\ ,\cr
S_{1+}S_{1+}\ket{\vec\gamma_1}&=
{\cal P}_4\ket{3 i a,iQ}\ .\cr}\eqno(6.17)
$$

     For a general level $\ell$ in the two-scalar $W_3$ string, we may
consider building physical states by acting on any of the six tachyons, with
momenta given in (6.3), with arbitrary integer powers of the four screening
operators, whose momenta are given in (6.2).  The condition given after eq.\
(6.9) is equivalent to the requirement that the conformal weight $\Delta$ of
the resulting exponential operator of the physical state should be an
integer; in fact it is related to the level number by $\ell=4-\Delta$.  By
requiring that $\ell$ be an integer, we find that physical states built from
screening operators can only arise at levels for which we may write
$$
\ell=\ft1{16}m^2 +\ft1{48}n^2 -\ft1{12}\ ,\eqno(6.18)
$$
where $m$ and $n$ are integers.  This excludes the possibility of having
such physical states at levels $\ell=7,\, 12,\, 17,\, 21,\, 22,\, 32,
\ldots$.  In fact, one arrives at exactly the same restriction (6.18) by
simply requiring that the on-shell momentum of a level-$\ell$ physical state
should be such that the its components $p$ and $\beta$ be integer multiples
of $\ft17 i a$ and $\ft17 i Q$ respectively.

\bigskip\bigskip
\noindent{\bf 7. Conclusions}
\bigskip\bigskip

     In this paper we have found the spectrum of the open $W_3$ string at
low levels.  For the case of the $(D+2)$-dimensional $W_3$ string, we
carried out the analysis up to and including level 2.  For the case of the
two-scalar $W_3$ string, we extended the analysis to level 4.  This was done
by solving the physical-state conditions at these levels and comparing the
results with the null states.  It was found that for the two-scalar $W_3$
string all the physical states except those at level 0 were null, and so at
least up to level 4 the theory contains no degrees of freedom.  For the
$(D+2)$-scalar $W_3$ string, however, it was found that any state that
contained oscillators of the field $\varphi^{(2)}$ was null.  The physical
states are contained in the remaining Fock space, $\tilde {\cal H}$
generated by the oscillators of the remaining $D+1$ fields subject to
$$
(\tilde L_0-a^{\rm eff})\ket{\psi}=0,\qquad \tilde L_n\ket{\psi}=0,\quad
n\ge 1,\eqno(7.1)
$$
where $\tilde L_0$ and $\tilde L_n$ are the Virasoro generators restricted
to the the oscillators in the $D+1$ fields, and $\ket{\psi}\in \tilde {\cal
H}$.  The intercept $a^{\rm eff}$ can take the values 1 or $\ft{15}{16}$,
leading to two sectors in the spectrum of physical states.  There exist
corresponding null states in each of these sectors, and the count of degrees
of freedom is given above.  For example, one finds at level zero two
tachyons, one from each sector.  At level 1, there are two vectors; one from
the $a^{\rm eff}=1$ sector with $D-1$ degrees of freedom, and the other from
the $a^{\rm eff}=\ft{15}{16}$ sector with $D$ degrees of freedom.  Although
the presence of background charges obscures the interpretation of mass, one
can take the above to mean that the open $W_3$ string describes only one
massless state, which is a ``photon.''  It is shown that at any level the
momentum in the $\varphi^{(2)}$ direction obeys a polynomial equation and at
the levels considered it has been found that the solutions are all integer
multiples of $\ft17 iQ$.  It seems likely that all these features generalise
to all levels.

     It was then shown, for the two-scalar $W_3$ string, that all the
physical states that were found could be described by the $W_3$ screening
operators acting on the vacuum.  This feature also seems likely to
generalise to all levels.   Although the closed $W_3$ string was not
considered in this paper, it is clear that it shares many of the above
features.  The only massless states in this case will be the graviton,
antisymmetric tensor, and scalar, just as for the usual closed bosonic
string.

\np
\noindent{\bf Note Added}
\bigskip

     After this work was completed, evidence for the existence of further
physical states in the $W_3$ string has been found.  As discussed in section
2, we have been concerned in this paper with physical states that have the
``standard'' form $\ket{\psi}\otimes \ket{\rm gh}$, where $\ket{\psi}$ is
built purely from matter oscillators acting on a momentum state, and
$\ket{\rm gh}$ is the ghost vacuum state.  In [17], it was shown that in
the two-scalar $W_3$ string there is a ``ground-ring'' structure of physical
states of ``non-standard'' form, involving ghost as well as matter
excitations.  These are analogous to the discrete states of the ordinary
two-dimensional string [18,19].  In fact, an example of a discrete state with
non-standard ghost structure was found in the two-scalar $W_3$ string in
[20].  In addition, it was found in [17] that there are continuous-momentum
physical states involving ghost as well as matter excitations in the
multi-scalar $W_3$ string too.  These physical states have no analogue in
ordinary string theory in more than two dimensions; their existence seems to
be related to the fact that the gauge symmetries of the states in the $W_3$
string are insufficient to permit the introduction of a physical gauge like
the light-cone gauge.  Further evidence for the existence of new physical
states was found in [21], where it was shown that there are poles in the
four-point scattering amplitude for the $\beta=iQ$ tachyons in (2.26) that
do not correspond to the masses of any of the physical states with
``standard'' ghost structure.  The details of these new physical states will
be discussed elsewhere [22].

\bigskip\bigskip

\centerline{\bf ACKNOWLEDGMENTS}
\bigskip

We are grateful to M. Freeman, K. Hornfeck, E. Sezgin and X.J. Wang for
discussions.  P.C.W. thanks the Newton Institute in Cambridge, and C.N.P. and
P.C.W. thank the Chalmers Institute of Technology in G\"oteborg, for
hospitality.

\np
\noindent{\bf Appendix A}
\bigskip
\bigskip

     Here we collect together some results for the operators $\tilde L_{-n}$
for the Virasoro algebra that act on a highest-weight state $\ket{h}$ with
weight $h$, and give a new highest-weight state $\tilde L_{-n}\ket{h}$, at
level $n$ and having weight $h+n$.  For each $n$ from 1 up to 12 we give the
set of permitted $h$ values, and, for convenience, the corresponding weights
$h+n$ of the resulting states.  The allowed values of $h$ depend, in
general, on the value $c$ of the central charge.  We present results here
for two values that are relevant for this paper, namely $c=26$ for the usual
bosonic string, and $c=25\ft12$ for the ``effective''  ordinary  sector of
the $W_3$ string.

     At $c=26$ we have:

\bigskip
\settabs 7 \columns

\+$\tilde L_{-1}:$ &$h=0$ &&&$\rightarrow$ &$1$ &\cr
\+$\tilde L_{-2}:$ &$h=-1,-\ft{13}8$ &&&$\rightarrow$ &$1,\ft38$ &\cr
\+$\tilde L_{-3}:$ &$h=-4,-\ft73$ &&&$\rightarrow$ &$-1,\ft23$ &\cr
\+$\tilde L_{-4}:$ &$h=-4,-\ft{25}8,-\ft{57}8$ &&&$\rightarrow$
& $0,\ft78,-\ft{25}8$ &\cr
\+$\tilde L_{-5}:$ &$h=-4,-6,-11$ &&&$\rightarrow$ &$1,-1,-6$ &\cr
\+$\tilde L_{-6}:$ &$h=-6,-\ft{25}3,-\ft{125}8,-\ft{119}{24}$
&&&$\rightarrow$ &$0,-\ft73,-\ft{77}8,\ft{25}{24}$ &\cr
\+$\tilde L_{-7}:$ &$ h=-6,-11,-21$ &&&$\rightarrow$ &$1,-4,-14$ &
$\qquad\qquad(A.1)$\cr
\+$\tilde L_{-8}:$ &$h=-14,-\ft{57}8,-\ft{77}8,-\ft{217}8$
&&&$\rightarrow$ &$-6,\ft78,-\ft{13}8,-\ft{153}8$ &\cr
\+$\tilde L_{-9}:$ &$h=-34,-\ft{25}3,-\ft{52}3$ &&&$\rightarrow$
&$-25,\ft23,-\ft{25}3$ &\cr
\+$\tilde L_{-10}:$ &$h=-11,-14,-21,-\ft{77}8,-\ft{333}8$ &&&$\rightarrow$
& $-1,-4,-11,\ft38,-\ft{253}8$ &\cr
\+$\tilde L_{-11}:$ &$h=-11,-25,-50$ &&&$\rightarrow$ &$0,-14,-39$ &\cr
\+$\tilde L_{-12}:$
&$h=-11,-\ft{88}3,-\ft{153}8,-\ft{473}8,-\ft{299}{24},-\ft{299}{24}$
&&&$\rightarrow$
&$1,-\ft{52}3,-\ft{57}8,-\ft{377}8,-\ft{11}{24},-\ft{11}{24}$ &\cr
\bigskip

We do not in general present the explicit forms for the $\tilde L_{-n}$
operators, since their structure is quite complicated.  The first few are
given below.  Note that the coefficients of the various terms in these
expressions will in general be $h$ dependent, as well as $c$ dependent.
One can see from the above results that $\tilde L_{-1}$, $\tilde L_{-2}$,
$\tilde L_{-5}$, $\tilde L_{-7}$ and $\tilde L_{-12}$ are relevant for
string theory, since they can produce highest-weight states with conformal
weight 1.  One can also see that the following relevant factorisations
occur:
$$
\eqalign{
\tilde L_{-5}(-4)&=\tilde L_{-1}\tilde L_{-4}=\tilde L_{-2}\tilde
L_{-3}\ ,\cr
\tilde L_{-7}(-6)&=\tilde L_{-1}\tilde L_{-6}=\tilde L_{-2}\tilde L_{-5}
\ ,\cr
\tilde L_{-12}(-11)&=\tilde L_{-1}\tilde L_{-11}=\tilde L_{-2}\tilde
L_{-10}\ .\cr}\eqno(A.2)
$$
Here, the conformal weight of the state on which the operator on the
left-hand side acts is indicated in brackets.  Presumably this structure of
factorisations persists for all higher $\tilde L_{-n}$ operators.

     The explicit forms of the first five $\tilde L_{-n}$ operators are
as follows.
$$
\tilde L_{-1} = L_{-1}\eqno(A.3)
$$
gives a null state if and only if it acts on a highest-weight state
of conformal weight $h=0$, independent of the value of $c$.  At level 2,
we have
$$
\tilde L_{-2} = L_{-2}-{3\over 2(2h+1)}L_{-1}^2\ ,\eqno(A.4)
$$
for which $h$ must be a root of the equation
$$
16h^2-10h + (2h+1)c=0\ .\eqno(A.5)
$$
At level 3,
$$
\tilde L_{-3} = L_{-3} -{2\over h} L_{-2} L_{-1} +{1\over h(h+1)} L_{-1}^3
\ ,\eqno(A.6)
$$
where $h$ must be a root of
$$
3h^2-7h+2+(h+1)c=0\ .\eqno(A.7)
$$
The level 4 operator has the form
$$
\tilde L_{-4} = \lambda_1 L_{-4} +\lambda_2 L_{-3} L_{-1} + \lambda_3
L_{-2}^2 + \lambda_4 L_{-2}L_{-1}^2 + \lambda_5 L_{-1}^4 \ ,\eqno(A.8)
$$
where the coefficients are given by
$$
\eqalign{
\lambda_1&= -\ft4{75}\big[16h^3+ (2c+6)h^2 + (c+50)h -3c+3 \big]
\lambda_5\ ,\cr
\lambda_2&= \ft2{15}\big[ 16h^2 +(2c+2)h + 3c+12\big]\lambda_5\ ,\cr
\lambda_3&= \ft4{45}\big[ 4h^2 +(-2c+38)h -3c+3\big]\lambda_5\ ,\cr
\lambda_4&= -\ft 43(2h+3)\lambda_5 \ .\cr}\eqno(A.9)
$$
The conformal weight of the state on which it acts must be a root of
$$
(8h+c-1)\big[16h^2+(10c-82)h +15c+66\big]=0\ .\eqno(A.10)
$$
Finally, at level 5 we have
$$
\tilde L_{-5}= \mu_1 L_{-5} + \mu_2 L_{-4} L_{-1} + \mu_3 L_{-3} L_{-2}
+\mu_4 L_{-3} L_{-1}^2 + \mu_5 L_{-2}^2 L_{-1} +\mu_6 L_{-2} L_{-1}^3 +
\mu_7 L_{-1}^5\ ,\eqno(A.11)
$$
where the coefficients are given by
$$
\eqalign{
\mu_1&= \ft19 h\big[5h^3 +(c+1)h^2 +(2c+24)h +12\big]\mu_7\ ,\cr
\mu_2&= -\ft13\big[5h^3+(c+1)h^2 +(2c+24)h +12\big]\mu_7\ ,\cr
\mu_3&= -\ft19h\big[7h^2+(-c+41)h-2c+18\big]\mu_7\ ,\cr
\mu_4&= \ft13\big[8h^2+(c+4)h +2c+12\big]\mu_7\ ,\cr
\mu_5&= \ft29\big[7h^2+(-c+41)h-2c+18\big]\mu_7\ ,\cr
\mu_6&= -\ft{10}3 (h+2)\mu_7\ .\cr}\eqno(A.12)
$$
In this case, the conformal weight of the state on which the operator acts
must satisfy the two equations
$$
\eqalign{
h(14h+c+30)\big[h^2 +(c-9)h +2c+14\big]&=0\ ,\cr
(13h+2c)\big[h^2+(c-9)h+2c+14\big]&=0\ .\cr}\eqno(A.13)
$$
Similar, but progressively more complicated, results can be obtained at all
higher levels too.  In each case, by solving the conditions that follow
from requiring that $L_1 \tilde L_{n-}\ket{h}=0$ and $L_2 \tilde
L_{-n}\ket{h}=0$, one can determine the various coefficients in the
expression for $\tilde L_{-n}$ and also the polynomial conditions relating
$c$ and $h$.  It is these values of $h$ that are tabulated in eq.\ $(A.1)$
for $c=26$, and in eq.\ $(A.14)$ below for $c=25\ft12$.

     Another case of interest here is when $c=25\ft12$, since this is the
value of the central charge for the ``effective'' Virasoro string theory of
the case 3-states in section 3.  The corresponding permitted values for $h$
are

\+$\tilde L_{-1}:$ &$h=0$ &&&$\rightarrow$ &$1$ &\cr
\+$\tilde L_{-2}:$ &$h=-\ft32,-\ft{17}{16}$ &&&$\rightarrow$
&$\ft12,\ft{15}{16}$ &\cr
\+$\tilde L_{-3}:$ &$h=-\ft52,-\ft{11}3$ &&&$\rightarrow$
&$\ft12,-\ft23$ &\cr
\+$\tilde L_{-4}:$ &$h=-\ft{13}2,-\ft{49}{16},-\ft{69}{16}$
&&&$\rightarrow$ & $-\ft52,\ft{15}{16},-\ft{5}{16}$ &\cr
\+$\tilde L_{-5}:$ &$h=-10,-\ft{13}2$ &&&$\rightarrow$ &$-5,-\ft32$ &\cr
\+$\tilde L_{-6}:$ &$h=-5,-\ft{85}6,-\ft{145}{16},-\ft{275}{48}$
&&&$\rightarrow$ &$1,-\ft{49}6,-\ft{49}{16},\ft{13}{48}$ &\cr
\+$\tilde L_{-7}:$ &$ h=-12,-19,-\ft{13}2$ &&&$\rightarrow$ &$-5,-12,\ft12$
&$\qquad\qquad(A.14)$\cr
\+$\tilde L_{-8}:$ &$h=-\ft{49}2,-\ft{117}{16},-\ft{145}{16},-\ft{245}{16}$
&&&$\rightarrow$ &$-\ft{33}2,\ft{11}{16},-\ft{17}{16},-\ft{117}{16}$ &\cr
\+$\tilde L_{-9}:$ &$h=-19,-\ft{92}3,-\ft{49}6$ &&&$\rightarrow$
&$-10,-\ft{65}3,\ft56$ &\cr
\+$\tilde L_{-10}:$
&$h=-10,-\ft{75}2,-\ft{145}{16},-\ft{209}{16},-\ft{369}{16}$ &&&$\rightarrow$
& $0,-\ft{55}2,\ft{15}{16},-\ft{49}{16},-\ft{209}{16}$ &\cr
\+$\tilde L_{-11}:$ &$h=-10,-45,-\ft{55}2$
&&&$\rightarrow$ &$1,-34,-\ft{33}2$ &\cr
\+$\tilde L_{-12}:$
&$h=-12,-\ft{319}6,-\ft{209}{16},-\ft{517}{16},-\ft{527}{48},-\ft{851}{48}$
&&&$\rightarrow$
&$0,-\ft{247}6,-\ft{17}{16},-\ft{325}{16},\ft{49}{48},-\ft{275}{48}$ &\cr
\bigskip

     For the effective string theory, two values of the conformal weight
are relevant, namely $L_0^{\rm eff}=1$ and $L_0^{\rm eff}=\ft{15}{16}$.  For
the first of these, we see that the following factorisation occurs:
$$
\tilde L_{-11}(-10)=\tilde L_{-1}\tilde L_{-10}=\tilde L_{-6}\tilde L_{-5}
\ .\eqno(A.15)
$$
For $L_0^{\rm eff}=\ft{15}{16}$, we see the factorisation
$$
\tilde L_{-10}(-\ft{145}{16})=\tilde L_{-2}\tilde L_{-8}=\tilde L_{-4}\tilde
L_{-6} \ .\eqno(A.16)
$$
These factorisations illustrate what is presumably a generic pattern, namely
that at $c=25\ft12$ the operators that give conformal-weight 1 effective
physical null states can all be factorised two ways, as either $\tilde
L_{-1}$ or $\tilde L_{-6}$ acting on lower-level operators; and the operators
that give conformal-weight $\ft{15}{16}$ physical null states can also all be
factorised two ways, as either $\tilde L_{-2}$ or $\tilde L_{-4}$ acting on
lower-level operators.  For comparison, at $c=26$, the operators that give
conformal-weight 1 physical null states can all be factorised two ways too,
as either $\tilde L_{-1}$ or $\tilde L_{-2}$ acting on lower-order
operators.  The significance of the level-numbers of the two
``fundamental'' factorising operators in each case can be understood by
looking at the level numbers at which physical null states occur in the
relevant 1-scalar string or effective string.  By using the screening-charge
analysis described in section 5, one easily sees that at $c=26$ such null
states occur at levels $\ell=\ft16n(n+1)=0,1,2,5,7,12, 15, 22\ldots$,
where $n$ (mod) 3 $=0,2$.  A similar analysis at $c=25\ft12$ shows that null
states in a 1-scalar effective string with $L_0^{\rm eff}=1$ occur at levels
$\ell=\ft1{12}n(n+1)=0,1,6,11,13, 20, 35\ldots$, where $n$ (mod) 12
$=0,3, 8, 11$.  In the case that $L_0^{\rm eff}=\ft{15}{16}$, the null states
occur at levels $\ell=\ft13 n(n+1)=0,2,4,10,14,24, 30\ldots$, where
$n$ (mod) 3 $=0,2$.  Thus in all cases, the fundamental factorising
operators occur at the lowest two non-zero level numbers.  It is a fluke of
ordinary string theory that these happen to be levels 1 and 2.

\np

\singlespace
\centerline{\bf REFERENCES}
\frenchspacing
\bigskip

\item{[1]}A.B. Zamolodchikov, {\sl Teo. Mat. Fiz.} {\bf 65} (1985) 347.

\item{[2]}J. Thierry-Mieg, {\sl Phys. Lett.} {\bf 197B} (1987) 368.

\item{[3]}A. Bilal and J.-L. Gervais, {\sl Nucl. Phys.} {\bf B314} (1989)
646.

\item{[4]}P.S. Howe and P.C. West, unpublished.

\item{[5]}C.N. Pope, L.J. Romans and K.S. Stelle, {\sl Phys.
Lett.} {\bf 268B} (1991) 167.

\item{[6]}C.N. Pope, L.J. Romans and K.S. Stelle, {\sl Phys. Lett.}
{\bf 269B} (1991) 287.

\item{[7]}V.A. Fateev and S. Lukyanov, {\sl Int. J. Mod.  Phys.} {\bf
A3} (1988) 507;\nl
V.A. Fateev and A. Zamolodchikov, {\sl Nucl.  Phys.}  {\bf B280} (1987) 644.

\item{[8]}L.J.  Romans, {\sl Nucl.  Phys.} {\bf B352} (1991) 829.

\item{[9]}S.R. Das, A. Dhar and S.K. Rama, {\sl Mod. Phys. Lett.}
{\bf A6} (1991) 3055; {\sl Int. J. Mod. Phys.} {\bf A7} (1992) 2295.

\item{[10]}C.N.\ Pope, L.J.\ Romans, E.\ Sezgin and K.S.\ Stelle,
{\sl Phys. Lett.} {\bf 274B} (1992) 298.

\item{[11]}H. Lu, C.N. Pope, S. Schrans and K.W. Xu,  {\sl Nucl.
Phys.} {\bf B385} (1992) 99.

\item{[12]}H. Lu, C.N. Pope, X.J. Wang and K.W. Xu, {\sl Nucl. Phys.} {\bf
B379} (1992) 24.

\item{[13]}E. Braaten, T. Curtright and C. Thorn, {\sl Ann. Phys.} {\bf 147}
(1983) 365.

\item{[14]}E. Braaten, T. Curtright, G. Ghandour and C. Thorn, {\sl Ann.
Phys.} {\bf 153} (1984) 147.

\item{[15]}V.G. Kac, ``Infinite-dimensional Lie Algebras,'' Cambridge
University Press, (1985).

\item{[16]}D. Friedan, Z. Qiu and S.H. Shenker, {\sl Comm. Math. Phys.} {\bf
107} (1986) 535.

\item{[17]}C.N. Pope, E. Sezgin, K.S. Stelle and X.J. Wang, ``Discrete
States in the $W_3$ String,'' preprint, CTP TAMU-64/92,
Imperial/TP/91-92/40, hep-th/9209111, to appear in Phys. Lett. B.

\item{[18]}A.M. Polyakov, {\sl Mod. Phys. Lett.} {\bf A6} (1991) 635.

\item{[19]}E. Witten, {\sl Nucl. Phys.} {\bf B373} (1992) 187.

\item{[20]}S.K.\ Rama, {\sl Mod.\ Phys.\ Lett.}\ {\bf A6} (1991) 3531.

\item{[21]}M.D. Freeman and P.C. West, ``$W_3$ String Scattering,''
preprint, KCL-TH-92-4.

\item{[22]}H. Lu, C.N. Pope and P.C. West, in progress.

\end